\documentclass[a4paper,12pt]{JHEP3}
\usepackage{graphicx}
\usepackage{epsfig}
\usepackage{amsmath, amssymb}

\def\dj{\hbox{d\kern-0.347em \vrule width 0.3em height 1.252ex depth
-1.21ex \kern 0.051em}}

\def\Re{{\rm Re\,}}

\def\ran{\rangle}
\def\lan{\langle}

\def\in{{\rm{in}}}

\newcommand{\ra}{\rightarrow}

\newcommand{\be}{\begin{equation}}
\newcommand{\ee}{\end{equation}}
\newcommand{\ben}{\begin{equation*}}
\newcommand{\een}{\end{equation*}}
\newcommand{\bea}{\begin{eqnarray}}
\newcommand{\eea}{\end{eqnarray}}
\newcommand{\bean}{\begin{eqnarray*}}
\newcommand{\eean}{\end{eqnarray*}}
\newcommand{\brr}{\begin{array}}
\newcommand{\err}{\end{array}}
\newcommand{\bc}{\begin{center}}
\newcommand{\ec}{\end{center}}
\newcommand{\nn}{\nonumber }

\newcommand{\lsim}{\,\raisebox{-0.6ex}{$\buildrel < \over \sim$}\,}
\newcommand{\gsim}{\,\raisebox{-0.6ex}{$\buildrel > \over \sim$}\,}

\newcommand{\bk}{{\mathbf k}}

\newcommand{\by}{{\mathbf y}}
\newcommand{\bx}{{\mathbf x}}

\newcommand{\bq}{{\mathbf q}}

\newcommand{\GG}{{\cal G}}
\newcommand{\HH}{{\cal H}}
\newcommand{\II}{{\cal I}}
\newcommand{\MM}{{\cal M}}
\newcommand{\NN}{{\cal N}}
\newcommand{\PP}{{\cal P}}

\newcommand{\cd}{\cdot}
\newcommand{\al}{\alpha}

\newcommand{\de}{\delta}

\newcommand{\ep}{\epsilon}

\newcommand{\Ga}{\Gamma}

\newcommand{\la}{\lambda}
\newcommand{\Om}{\Omega}

\newcommand{\lp}{\left}
\newcommand{\rp}{\right}
\newcommand{\ti}{\tilde}
\newcommand{\cf}{\emph{c.f.}}

\title{Can the observed large scale magnetic fields be seeded by helical 
       primordial fields?}

\author{Chiara Caprini\\ IPhT, CEA-Saclay, CNRS, URA 2306, F-91191 Gif-sur-Yvette, France
 \\
\email{chiara.caprini@cea.fr}}
\author{Ruth Durrer\\  D\'epartement de Physique Th\'eorique, Universit\'e de
  Gen\`eve, 24 quai Ernest Ansermet, CH--1211 Gen\`eve 4, Switzerland\\
\email{ruth.durrer@unige.ch}}
\author{Elisa Fenu\\
D\'epartement de Physique Th\'eorique, Universit\'e de
  Gen\`eve, 24 quai Ernest Ansermet, CH--1211 Gen\`eve 4, Switzerland\\
  \email{elisa.fenu@unige.ch}}

\received{\today}

\keywords{Cosmology, Magnetic fields, Helicity, Gravitational Waves}

\abstract{Gravitational wave production induces a strong constraint on the amplitude of a 
primordial magnetic field. It has been shown that the 
nucleosynthesis bound for a stochastic gravitational wave background 
implies that causally generated fields cannot have enough power on large scales to provide 
the seeds necessary for the observed magnetic fields in 
galaxies and clusters, even by the most optimistic dynamo amplification. Magnetic fields generated at inflation can have high enough amplitude only if their spectrum is very red. Here we show that helicity, which leads to an inverse cascade, can 
mitigate these limits. In particular, we find that helical fields generated at the QCD phase transition or at inflation with red spectrum are possible seeds for the dynamo. Helical fields generated at the electroweak phase transition are instead excluded as seeds at large scales. 
We also calculate the spectrum of gravitational waves generated by helical magnetic fields.}

\begin{document}

\section{Introduction}
Magnetic fields are ubiquitous in the Universe. Wherever they can be 
measured, they are found. In stars, in galaxies~\cite{Bgal}, locally and at 
high redshift~\cite{Bhiz}, and in clusters of galaxies~\cite{Bclust}. 
There is also evidence of magnetic fields in super clusters~\cite{kron.new}.
However, the origin of these fields is still unclear. Have they emerged in the late Universe from 
charge separation processes or by ejection from stars and galaxies~\cite{Subramanian:2008tt}? Or have they been amplified 
from primordial seed fields which may represent a relic from the early Universe,
from the electroweak (EW) phase transition~\cite{ew} or even from 
inflation~\cite{infla}? If the second exciting possibility is realized, this 
means that we can learn about processes in the early universe from studying 
cosmological large scale magnetic fields.

In a previous paper~\cite{Caprini:2001nb} it has been shown that 
primordial magnetic fields lead to significant production of gravitational 
waves. If the magnetic field spectrum is blue, as it has to be if the 
production mechanism is causal~\cite{causal}, the nucleosynthesis limit for 
a gravitational wave (GW) background strongly constrains the amplitude of magnetic 
fields on large scales. This strong constraint comes from the fact that for 
causal magnetic fields, the energy density has to behave like
\be \frac{d\rho_B(k)}{d\log k} \propto k^5 \ee
with comoving wave number $k$, on scales which are larger than the 
correlation scale. Hence even the moderate nucleosynthesis limit, since it 
comes from the smallest scales, highest wave numbers, at which the magnetic 
field is maximal, leads to a very strong limit on the field amplitude at 
large, cosmological scales. The detailed results are given 
in~\cite{Caprini:2001nb,Caprini:2006jb}. For the   derivation of this limit it is assumed 
that the magnetic field spectrum evolves  
solely via the damping of fields on small scales and via flux conservation. 
On large scales, the 
magnetic field spectrum scaled to today is assumed to remain constant.

However, if the magnetic field has non-vanishing helicity, the 
conservation of helicity leads to an inverse cascade, {\it i.e.} it can move power 
from small to large scales.  A derivation of this result can be 
found  in the review~\cite{BS}. This can 
 mitigate the magnetic field limit 
which precisely comes from the fact that for causally produced magnetic 
fields there is so little power on large scales. The production of helical 
magnetic fields has been proposed for both, inflation~\cite{infla:hel} and the EW phase transition where the magnetic 
field helicity is linked to the baryon number~\cite{ew:hel}. Furthermore,
the formation of maximally helical magnetic fields at the QCD phase 
transition has been proposed in Ref.~\cite{QCD}.

The evolution of helical magnetic fields and the inverse cascade have been
studied in numerical simulations, and simple fits which describe the
evolution of the correlation scale $L(t)$ and of the magnetic field energy 
density $\rho_B(t)$ have been derived in Refs.~\cite{Christensson:2000sp,Banerjee:2004df,Campanelli:2007tc}.
Using these results, we want to determine upper bounds on the amplitude of {\em helical} magnetic fields from the induced
GWs. These bounds are summarised in Table \ref{tab}. 

In Section~\ref{s:setup} we present the basic definitions and discuss the 
evolution of normal and helical magnetic fields. Here we make use of the 
results for the inverse cascade discussed in Ref.~\cite{Campanelli:2007tc}. In  
Section~\ref{s:spec} we calculate the induced GW spectrum. 
In Section~\ref{s:res} we derive the limits on helical magnetic fields on 
cosmological scales.  In Section~\ref{s:con} we conclude.

{\bf Notation:} Throughout this paper we neglect curvature and the 
cosmological constant, which are not relevant for our discussion. The 
cosmological metric is given by
\be
ds^2 = a^2(t)\left( -dt^2 + \de_{ij}dx^idx^j\right)\,,
\ee
where $t$ denotes conformal time and the scale factor, normalized to $1$ 
today, is given to a good approximation by
\be
a(t) \simeq H_0t\lp[ \frac{H_0t}{4}+\sqrt{\Omega_{\rm rad}}\lp( \frac{g_0}{g_{\rm eff}(t)}\rp)^{1/6} \rp] ~,
\ee
where $H_0$ denotes the present value of the Hubble parameter, $g_{\rm eff}(t)$ is the number of effective relativistic degrees of freedom at time $t$, 
$g_0\equiv g_{\rm eff}(t_0) =2$, and $\Om_{\rm rad}$ is the radiation density parameter today. In the following, the density parameter is defined as $\Om_X(t) = \rho_X(t)/\rho_c(t)$, where $\rho_c(t)$ denotes the critical energy density at time $t$.

Spatial vectors are indicated in bold face, 3d spatial indices are lower 
case Latin letters while 4d spacetime indices are lower case Greek letters.


\section{The evolution of helical magnetic fields}\label{s:setup}
\subsection{Basic definitions}
The high conductivity of the cosmic plasma implies that, to lowest order,
magnetic fields evolve by flux conservation, so that $B\propto a^{-2}$.
We are mainly interested in the part of the time dependence of
our quantities which is not simply due to the expansion of the
Universe but to the growth of the magnetic correlation length and to the
additional decay of the magnetic energy density due to dissipation and to the MHD 
cascade~\cite{Christensson:2000sp,Banerjee:2004df,Campanelli:2007tc}. 
Therefore we eliminate the scaling with redshift by
expressing all the quantities in terms of comoving ones scaled to today
which we denote by a tilde. For example the comoving
magnetic energy density  is given by \cite{Caprini:2006jb}
\be
  \lan \ti B^2(t) \ran=  \lan  B^2(t) \ran a^4(t)  ~.
\ee
Here $\lan \ti B^2(t) \ran$  depends on time via the evolution of the
correlation length and because of energy dissipation.
Comoving quantities are not multiplied by powers of the 
scale factor $a^2(t)$ when lowering and rising indices.

The power spectrum of the magnetic field can be written as~\cite{Caprini:2003vc}
\be
  \lan \ti B_i(\bk,t) \ti B^*_j(\bq,t) \ran = \frac{(2\pi)^3}{2}
       \delta^3(\bk-\bq) \lp[\lp(\delta_{ij}-\hat{k}_i \hat{k}_j \rp)
       S(k,t) +   i\epsilon_{ijn}\hat{k}_n A(k,t)\rp]
       ~.  \label{e:Bansatz}
 \ee

The functions $S$ and $A$ denote the parity even and the parity odd
parts of the two point correlator respectively and $\bk$ is the
comoving wave vector (on the quantities where there is no danger of confusion
because they always denote conformal quantities like $\bk$, $S$ or $A$,
we omit the tilde). 
Using the above expressions, we compute the comoving magnetic energy density
$\tilde\rho_B =(8\pi)^{-1}\tilde B^2$,
\be  \label{e:rhoB}  
   \ti\rho_B(t) =\int_0^\infty \frac{ d k}{k} 
      \frac{ d\ti\rho_B(k,t)}{ d\log k} ~, \qquad 
   \frac{ d\ti\rho_B(k,t)}{d \log k} = \frac{k^3 S(k,t)}
      {2(2\pi)^3} ~.
\ee
Note that the antisymmetric part of the spectrum does
not contribute to the energy density but its presence, which indicates 
non-vanishing helicity,
influences the time dependence of $\ti\rho_B$ and of the magnetic correlation
length. In~\cite{Campanelli:2007tc} the author derives analytically the expressions for
the time evolution of the magnetic energy density and of the magnetic 
correlation length. These evolution laws have also been obtained in numerical simulations of a magnetic field in a
turbulent MHD phase during the radiation dominated era \cite{Banerjee:2004df}. On the other hand, the simulations of Ref.~\cite{Christensson:2000sp} indicate different exponents for the evolution laws. In the following analysis we adopt the analytical picture and the evolution laws derived in~\cite{Campanelli:2007tc}. It is easy to show that 
during this epoch the MHD equations are identical to those of a non-expanding 
Universe, provided that all physical variables are replaced by comoving 
variables (see~\cite{Lemoine,Banerjee:2004df}).

Our aim is to compute the GWs generated by an helical magnetic 
field and to use the GW energy density to derive constraints 
on the magnetic field strength on the comoving scale 
$\tilde\la \simeq 0.1$ Mpc. For this we express the results of 
Ref.~\cite{Campanelli:2007tc} in terms of the quantities introduced above. 
Campanelli~\cite{Campanelli:2007tc} defines the spectral energy density 
$\varepsilon_B(k,t)$ which is related to Eq.~(\ref{e:rhoB}) 
by\footnote{Ref.~\cite{Campanelli:2007tc} is using Heavyside-Lorentz units 
such that $\rho_B =(1/2)B^2$ while we are using Gaussian units with 
$\rho_B =(1/8\pi)B^2$. This leads to differences of factors of $4\pi$ in the 
relative expressions.}
\be\label{e:def:varep}
  \frac{ d\ti\rho_B(k,t)}{ d \log k} =\frac{k^3 S(k,t)}
      {2(2\pi)^3} = k \, \varepsilon_B(k,t)   ~.
\ee
We adopt the following power spectrum for the magnetic field energy:
\be
  \label{epsK}
  \varepsilon_B(k,t) = \left\{ \begin{array}{ll}
\eta_B(t) \frac{\ti L(t) K^{n+2}}{(1+K^2)^{(7+2n)/4}} & {\rm for } \quad 
   K\leq \frac{ \ti L(t)}{\ti l_{\rm diss}(t)}\,, \vspace{0.1cm} \\
 0 &  {\rm for } \quad K\ge \frac{ \ti L(t)}{\ti l_{\rm diss}(t)} \,.
\end{array} \right.
\ee
Here $K\equiv k \ti L(t)/(2\pi) $, $\ti L(t)$ is the time dependent comoving
correlation length which we infer from~\cite{Campanelli:2007tc} and 
$\ti l_{\rm diss}$  is the comoving dissipation scale (for smaller scales 
the magnetic power spectrum is exponentially suppressed and we thus can set it 
to zero). We derive the dissipation scale $\ti l_{\rm diss}(t)$ in 
Appendix~\ref{A:ldiss}. Eq.~(\ref{epsK}) is inspired by Eq.~(19) of \cite{vonKarman}, which provides a continuous expression for the turbulent velocity  spectrum, interpolating between the large and small scale behaviours. We extend it here to the case of a magnetic field processed by MHD turbulence.  The initial power spectrum at large scales, $k\ti L\ll 1$ behaves 
like $k^2S(k) \propto k^{n+2}$, $n$ is the spectral index introduced in
Refs.~\cite{Caprini:2006jb,Caprini:2001nb}. If the initial 
correlation length is finite, the power spectrum given in 
Eq.~(\ref{e:Bansatz}) must be analytic and hence $n\ge 2$ must be
an even integer~\cite{causal}. A scale invariant spectrum corresponds 
to $n=-3$. We have chosen the form of
$\varepsilon_B(k,t)$ such that it scales like $k^{-3/2}$ in the inertial range,
$2\pi/\ti L(t) < k < 2\pi/ \ti l_{\rm diss}$. This behavior corresponds to the 
Iroshnikov--Kraichnan spectrum, which can be realised in 
fully developed MHD turbulence~\cite{Shore}. Here we could have equally chosen the Kolmogorov or Goldreich-Sridhar spectral slopes: our main result does not depend significantly on this choice. Note that Ref.~\cite{Campanelli:2007tc} does not account for the presence of the inertial range. 

We adopt here the interpolating formula (\ref{epsK}) in order to avoid joining the two asymptotic behaviours, at large and small scales, of the magnetic field spectrum up to the energy injection scale $\ti L(t)$. This has been done for example in Ref.~\cite{Caprini:2001nb}, and it leads to an overestimation of the peak amplitude. Note that we extend the formula of Ref.~\cite{vonKarman} to represent also magnetic fields with red spectra, generated during inflation. 
  
Integrating Eq.~(\ref{epsK}) over $k$ we obtain the total comoving magnetic 
field energy density,
\be\label{normalizationK}
\ti\rho_B(t) = \rho_c(t_0)\ti\Om_B(t) \simeq \eta_B(t)\frac{\pi
 \Ga\left(\frac{1}{4}\right)\Ga\left(\frac{n+3}{2}\right)}{\Ga
\left(\frac{2n+7}{4}\right)}\,,
\ee
where we have set the upper limit of integration to infinity.
Hence the function $\eta_B(t)$ reflects the time dependence of the total energy density.
In the above expression we have introduced
\bea
\ti\Om_B(t) &\equiv& \frac{\ti\rho_B(t)}{\rho_c(t_0)} = \frac{\rho_B(t)a^4(t)}{\rho_c(t_0)} = 
 \frac{\rho_B(t)a^4(t)}{\rho_{\rm rad}(t_0)} \, \Om_{\rm rad} 
\nonumber \\  \label{Omega_B} 
 &=& \left(\frac{g_0}{g_{\rm eff}(t)}\right)^{1/3} 
   \frac{\rho_B(t)}{\rho_{\rm rad}(t)} \, \Om_{\rm rad}
 =  \left(\frac{g_0}{g_{\rm eff}(t)}\right)^{1/3} \Om_B(t) \, 
    \Om_{\rm rad} \,.
\eea
We restrict to the analysis of magnetic fields in a radiation dominated 
universe. Furthermore, adiabatic expansion implies~\cite{mybook} that the 
entropy $S \propto g_{\rm eff}(aT)^3$ is independent of time, so that 
$\rho_{\rm rad}(t)=\rho_{\rm rad}(t_0)a(t)^{-4} 
[g_0/g_{\rm eff}(t)]^{1/3}$. At time $t_*$, 
which we define as the  time at which turbulence is  fully developed (as we 
shall see in the following), one has
\be\label{Omega_B*}
\Om_B^* \equiv \Om_B(t_*) =\frac{\ti\Om_B(t_*)}{\Om_{\rm rad}} 
 \left(\frac{g_*}{g_0}\right)^{1/3}  \simeq 
\frac{\ti\Om_B(t_*)}{\Om_{\rm rad}} \,,
\ee
where $g_*\equiv g_{\rm eff}(t_*)$. The comoving magnetic energy density 
parameter $\ti \Om_{B}(t)$ depends on time via the dissipation of magnetic 
energy. 

In the following we will often neglect changes in  $g_{\rm eff}(t)$. For example, we neglect the fact that 
the evolution of gravitational radiation is modified, even if the universe is radiation 
dominated, due to the fact that while 
$g_{\rm eff}(t)$ is evolving the scale factor does not expand
like $a\propto t$ but somewhat faster. For the EW phase transition $(g_*/g_0)^{1/3} \simeq 3.7$.
In the magnetic field limits this factor enters at most with power $1/2$, 
which yields differences of less than a factor of two. In the amplitude
of gravitational waves it can enter with a higher power and change it by 
up to an order of magnitude. But this is in any case roughly the precision 
of the results derived in this work.

\subsection{Direct and inverse cascades}
\label{s:dirandinv}

The main difference between non helical field evolution (which 
only exhibits direct cascade and dissipation on small scales) 
and helical field evolution (which leads to inverse cascade) can be expressed 
in the time evolution of the correlation scale $\ti L(t)$ and the 
comoving magnetic field energy density $\tilde{\rho}_B(t)$, which we cast in the amplitude $\eta_B(t)$ \cite{Campanelli:2007tc}.

We introduce the normalized conformal time $\tau$
\be
  \tau = \frac{t-t_{\rm in}}{t_L^*}  ~,
\label{tau}
\ee
where $t_L^*$ denotes the initial eddy turn-over time on the scale of energy 
injection $\tilde{L}_*$, and $t_{\rm in}$ is the time at which the magnetic field is generated. 
The eddy turnover time is defined as $t_L^* \simeq \ti L_*/(2v_L)$, where $v_L$ is the initial eddy turnover speed and 
the factor $1/2$ comes from the fact the $\ti L_*$ is the eddy diameter. 

Starting from the asymptotic laws given in \cite{Campanelli:2007tc}, we rewrite 
the time evolution of the comoving magnetic energy density and of the comoving magnetic 
correlation length in terms of the normalised conformal time (\ref{tau}). However, with respect to \cite{Campanelli:2007tc} we add a model of the initial evolution in which the magnetic energy density grows continuously from zero to the equipartition value, at which MHD is fully developed. We do this because later on we evaluate the GW spectrum generated by the magnetic source, and the time continuity of the source does affect the resulting GW spectrum (see Ref.~\cite{CRTG} and section \ref{s:peak}). Therefore, we assume that the magnetic field energy density is zero 
(continuous) at $t=t_{\rm in}$, $\tau=0$; it then reaches its maximal value 
    $\ti\rho_B^*$ after a 'switching on' time which we set equal to the 
 characteristic eddy turnover time, $t_L^*$, namely at $t_*=t_{\rm in} + t_L^*$. We therefore define $t_* \equiv t_{\rm in}+t_L^*$ as the time when turbulence is fully developed, and normalise also the energy injection scale as $\ti L_*=\ti L(t_*)$, $\ti L(t_{\rm in})=0$. 
\begin{itemize}
\item
	Selective decay (direct cascade): the evolution laws are in this case
\bea  \label{e:EBsd}
	&& 	 \ti\rho_B(t) = \ti\rho_B^*
		 \left\{ \begin{array}{ll}
		 \tau & 
        \qquad \mbox{for } t_{\rm in} \leq t< t_* \,, \quad \tau \leq 1 \,,  \\
	 		\tau^{-2(n+3)/(n+5)}    & 	
          \qquad \mbox{for  }  t \geq t_*  \,, \quad \tau\geq 1 \,,
      			\end{array} \right.
	\\    \label{e:Lsd}
	  && \ti L(t) =\ti L_* \tau^{2/(n+5)}  ~,
	\eea
	where we may only consider $n>-2$ in order to recover the correct 
        behavior with respect to time of the above quantities ({\it i.e.} decay for the energy and growth for the correlation scale).
	The above expressions go continuously to zero with $\tau\ra 0$
        and they have the 
        asymptotic behavior found in Ref.~\cite{Campanelli:2007tc}  for 
        $t \gg t_{\rm in} \geq t_L^*$, $\tau\gg 1$. Note that 
 Ref.~\cite{Campanelli:2007tc} uses a spectral index $p$ which is related 
to $n$ via $p=n+2$. For simplicity, the parameter $\kappa_{\rm diss}$ of \cite{Campanelli:2007tc} is set to one, corresponding to $\gamma=3\Gamma^2[(1+p)/2]/(3+p)/\Gamma^2[p/2]$. \\

	The energy injection scale $\ti L_*$ is determined by the physical process that generates the magnetic field and the turbulence.
	Generically it can be parametrised as a fraction of the horizon at the initial time. 
       Therefore, we introduce the small parameter 
	$\epsilon< 1$ defined by
	 \be
	  \ti L_* = \epsilon \,t_\in ~,  \mbox{ such that }\quad    
	  t_L^* \simeq \frac{\ti L_*}{2v_L} = 
         \frac{\epsilon}{2v_L} t_\in  ~.  
	 \ee
         A typical value for causally generated turbulence is $\ep \simeq 10^{-2}$ (see for example~\cite{Caprini:2006jb}).
The necessary condition to have a turbulent cascade is that $t_L^*\leq t_{\rm in}$, {\it i.e.}
        $v_L\geq \epsilon/2$. Eddies of the size of the horizon which 
        move at the speed of light are the limiting case. 
	Although it grows, the correlation length 
	never becomes larger than the horizon.
	In fact one has 
	\be
		\frac{\ti L(t)}{t} = \lp[ \lp (2v_L\rp)^2 \ep^{n+3}\rp]^{1/(n+5)} \lp(1-\frac{t_\in}{t}\rp)^{2/(n+5)}
		 		\lp(\frac{t_\in}{t}\rp)^{(n+3)/(n+5)}  ~,
	\ee
	which is smaller than one for all times $t\geq t_{\rm in}$. Indeed, for the initial period $t_\in\leq t\leq t_*$, the term $ (1-t_\in/t)^{2/(n+5)} \to 0$ and it dominates the above expression, while for $t\gg t_*>t_\in$ the asymptotic behaviour is controlled
	by the last term of the equation which keeps the correlation length smaller than the Hubble radius. 
This is shown in Fig.~\ref{fig:lcorr}.	

\FIGURE[ht]{ 
	\epsfig{width=10cm, file=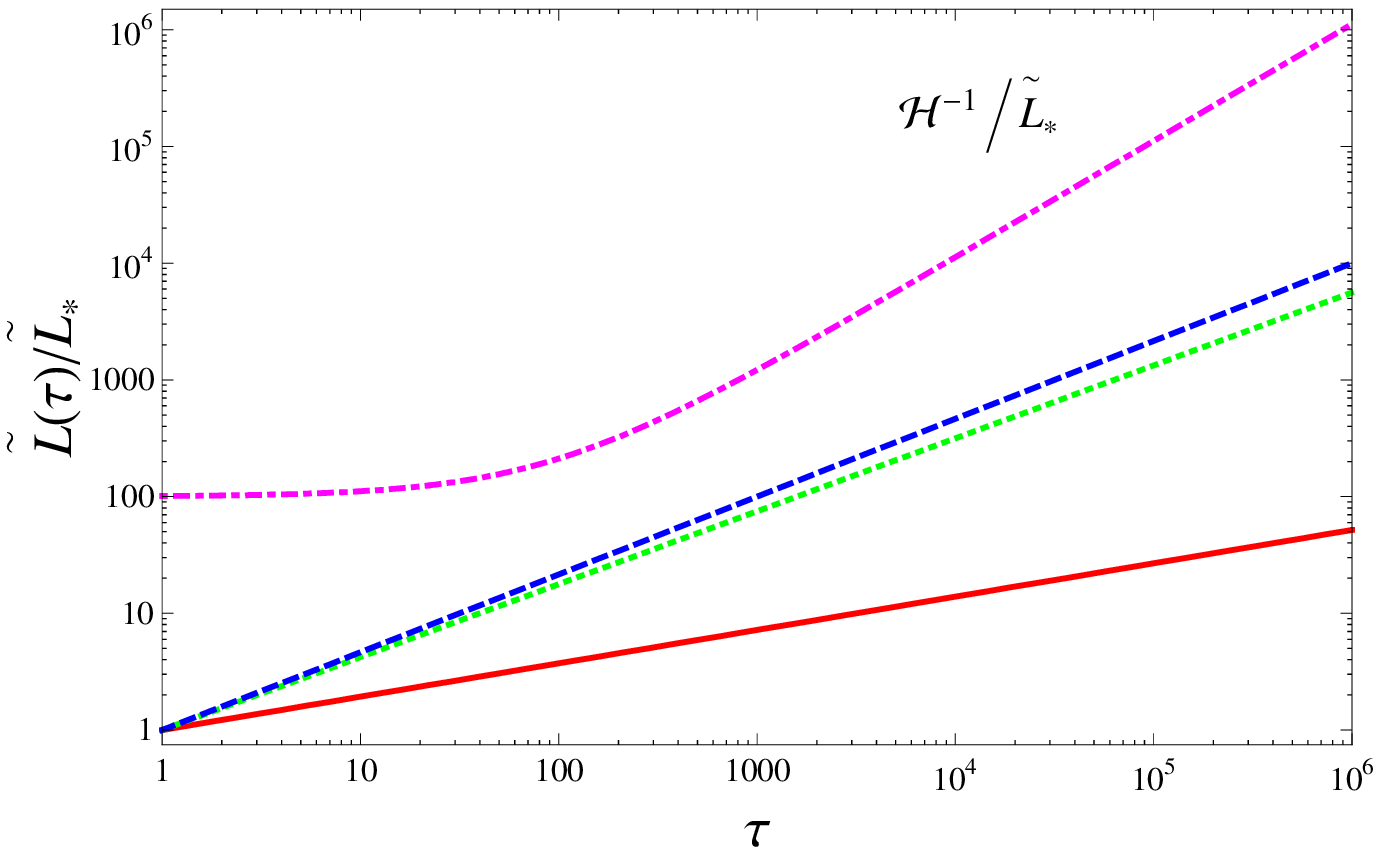}
\caption{Time evolution of the comoving correlation length $\ti L(\tau)$ as a 
function of $\tau$ for an inverse cascade (blue, dashed line) and for a direct cascade with the spectral indexes $n=2$ (red, solid 
line) and $n=-1.8$ (green, dotted line). They are compared with the time evolution of the comoving 
Hubble radius in a radiation dominated background, in units of the correlation
length $\ti L_*$ (magenta, dot-dashed line). }\label{fig:lcorr}}    

Even though the correlation length is growing,  the  spectral energy on a given comoving scale $k$ is at best constant. 
        On scales which are larger than the correlation scale, $k\ti L(t)<1$ 
 \be \frac{d\tilde\rho_B(k,t)}{d\log k} \propto  \ti\rho_B(t)(k\ti L(t))^{n+3} 
       \,. \label{drhoBdk}\ee
	From Eqs.~(\ref{e:EBsd}) and (\ref{e:Lsd}) it follows that during 
         selective 
        decay $\ti L^{n+3}(t) \ti\rho_B(t) =$ constant, hence 
        $d\tilde\rho_B / d\log k$ 
        does not evolve on large scales. The same behaviour is observed in the free decay of the turbulent velocity field, and is related to the constancy in time of Loitsyansky invariant (see for example \cite{Davidson,LL})
        
\item
	Inverse cascade: \\
	During inverse cascade we assume that the magnetic field energy and correlation 
        length evolve according to~\cite{Campanelli:2007tc} 
        \bea
        \label{Einverse}
		&&	\ti\rho_B(t) = \ti\rho_B^*
			\left\{ \begin{array}{ll} 
      	\tau     & 	\qquad \mbox{for } 
         t_{\rm in} \leq t< t_*\,, \quad \tau\leq 1\,,  \\
	        \tau^{-2/3}    & 	\qquad \mbox{for  }  
              t \geq t_* \,, \quad \tau\geq 1\,,
      			\end{array} \right.
	\\			
	\label{Linverse}
        && 	\tilde{L}(t) =h_B \tilde{L}_* \tau^{2/3}  ,
	\eea
	where $h_B$ is the initial fractional helicity: $h_B=0$ corresponds 
        to a non-helical magnetic field that remains non-helical for all its 
        evolution (for which the above scaling relations do not apply), 
    while $h_B=1$ characterizes a maximally helical field. The above equations
     are again valid only for $p=n+2> 0.$\\

	\FIGURE[ht]{ 
	\epsfig{width=10cm, file=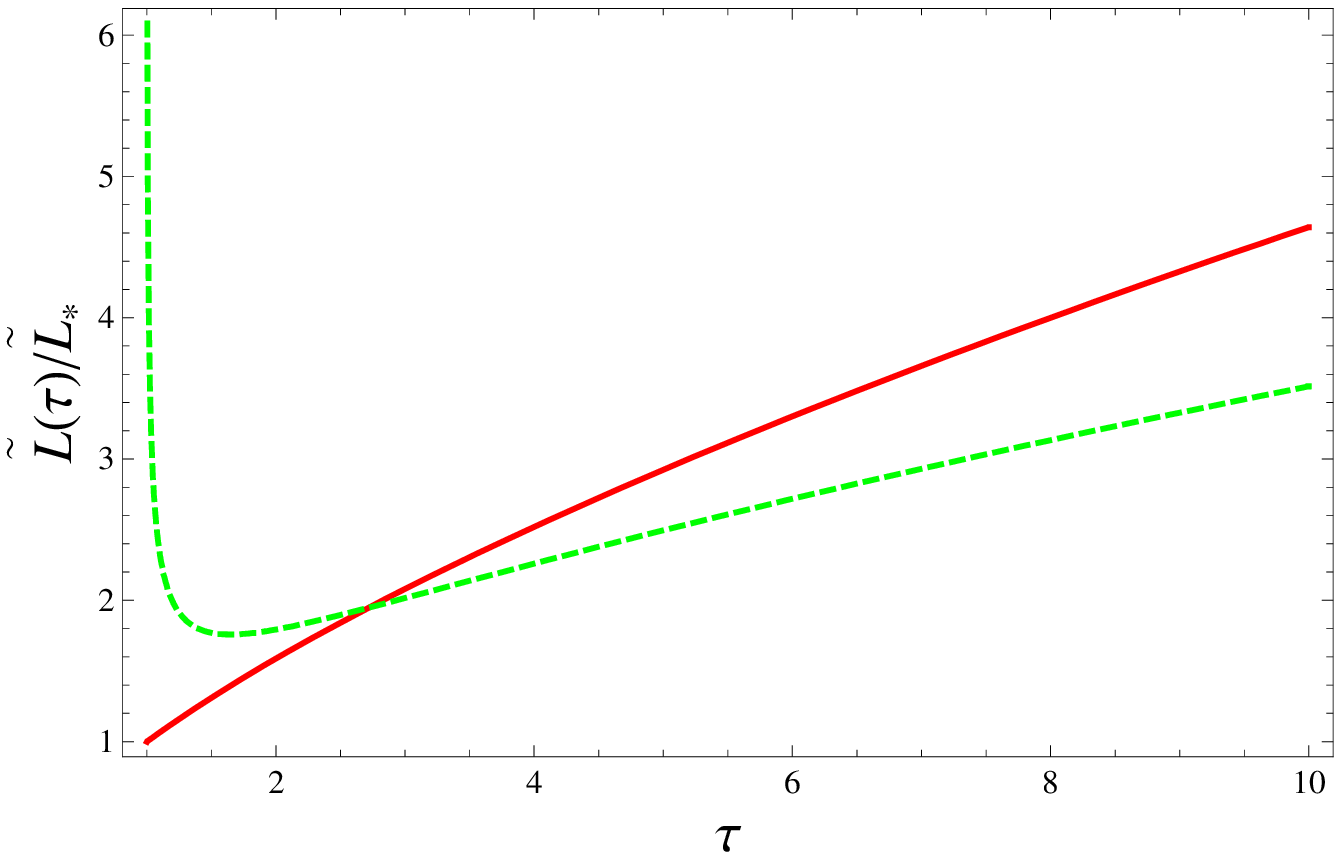}
	\caption{Time evolution of the comoving correlation length $\ti L(\tau)$ as a function of $\tau$ during
       inverse cascade as given in Ref.~\cite{Campanelli:2007tc} (green dashed line) and neglecting the logarithmic terms as in 
       Eq.~(\ref{Linverse}) (red solid line).        }
	\label{fig:Linverse}
	}

    The original expression of Ref.~\cite{Campanelli:2007tc} for $\tilde{L}(t)$ 
        contains also a logarithmic term, and gives asymptotically a 
        slower growth than Eq.~(\ref{Linverse}), as can be seen in Fig.~\ref{fig:Linverse}. In the following we neglect this logarithmic correction.
        Similarly, $\ti\rho_B(t)$ given by Eq.~(\ref{Einverse}) decays more rapidly than the 
	full expression given in~\cite{Campanelli:2007tc}, due to the same logarithmic correction (see Fig.~\ref{fig:Einverse}). 
         The limits on the magnetic field on large scales
       obtained using Eqs.~(\ref{Linverse}) and (\ref{Einverse})
	are less stringent than those one would obtain using the 
        more accurate expression of~\cite{Campanelli:2007tc}. Neglecting the logarithmic corrections is therefore a conservative assumption.

	\FIGURE[ht]{ 
	\epsfig{width=10cm, file=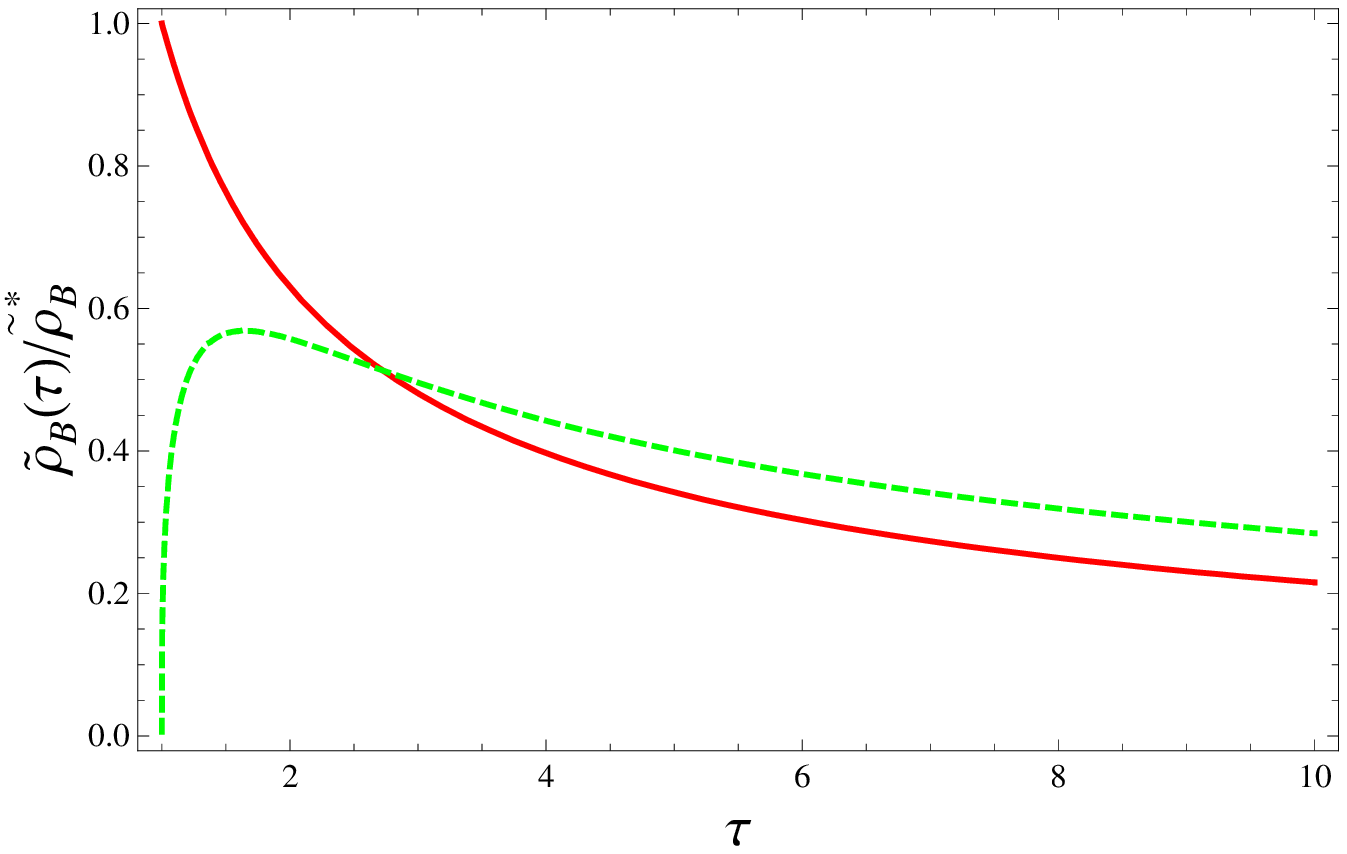}
	\caption{Time evolution of the magnetic energy density as a function of $\tau$ during inverse 
       cascade as given in Ref.~\cite{Campanelli:2007tc} (green dashed line)
        and neglecting the logarithmic terms as in Eq.~(\ref{Einverse}) (red solid line). 
        }
	\label{fig:Einverse}
	}       

        Although the correlation length grows faster than in the selective 
        decay phase, it never  becomes larger than the horizon even in this case (see Fig.~\ref{fig:lcorr}). 
	During the inverse cascade the product
        $\ti L(t) \ti\rho_B(t) =\ti L_* \ti\rho_B^*$ is constant. From Eq.~(\ref{drhoBdk}) one sees that during the inverse cascade 
	the spectral energy density is growing like $\ti L^{n+2}(t)$ at large scales.
\end{itemize}	

Eqs.~(\ref{Einverse}),~(\ref{Linverse}) 
apply only after a first phase during which the
system behaves as if the magnetic helicity was zero, {\it i.e.} by selective decay~\cite{Campanelli:2007tc}. Only when the magnetic field (with 
initial fractional helicity $h_B$) becomes maximally helical, the
inverse cascade can start, and Eqs.~(\ref{Einverse}) and
(\ref{Linverse}) apply.
In order to find the time at which this happens, one 
matches the product $\ti L(t)\ti\rho_B(t)$ (which is time dependent during  
selective decay), to its constant value during the inverse cascade.
This defines the transition time $t_h$
\be
  t_h = t_{\rm in}  \lp[1+ \frac{\epsilon}{2v_L} 
                                              h_B^{-(n+5)/(2n+4)} \rp]   ~.
\label{th}
\ee
For a maximally helical magnetic field $t_h(h_B=1) = t_*$.
In general, for a given $h_B$, the second stage takes place for  times 
$t>t_h\geq t_*$ and lasts until the  time $t_{\rm fin}$ at which the turbulent phase 
ends ({\it c.f.} next section). Moreover, in the case of zero initial helicity, one has pure direct cascade: $t_h \to \infty$ when $h_B\to 0$ 
(this is true only if we restrict the value of the spectral index to be $n>-2$, which we always do in the following). 

\subsection{The end of the turbulent phase and the dissipation scale}

The turbulent phase ends when the Reynolds number on the  scale of energy 
injection, $\ti L(t)$, becomes of order unity~\cite{LL}. In 
Appendix~\ref{A:Tfin} we calculate the epoch at which turbulence ends for the EW and the 
QCD phase transitions, as well as for inflation with $T_*\sim 10^{14}$ GeV.
The most important result from this calculation is that in all cases 
turbulence lasts for many Hubble times and therefore the source is not 
short lived. This finding and its consequences are the subject of
 \cite{CDS2}. For example, for a maximally helical field generated at the EW 
phase transition we find the final temperature $T_{\rm fin}\simeq 21\,$MeV 
(note that turbulence ends before nucleosynthesis \cite{Banerjee:2004df}).

In Appendix~\ref{A:ldiss} we determine the dissipation scale which is
the scale $\ti l_{\rm diss}(t)$ below which energy injection no longer leads to 
turbulence but is simply dissipated. This scale determines the 
time-dependent UV cutoff of our spectra. The evolution of both the correlation length $\ti L(T)$ and the 
dissipation scale $\ti l_{\rm diss}(T)$ for the EW phase transition is shown in 
Fig.~\ref{fig:ldiss}. The dissipation
scale grows faster than the correlation length and turbulence terminates 
roughly when the two curves cross. 

In the evaluation of these scales, we often use the approximation $T_1/T_2\simeq t_2/t_1$, which neglects changes in the number of effective relativistic degrees of freedom. Moreover in the following we do not distinguish among the temperatures corresponding to the initial time $t_{\rm in}$ and to the time at which turbulence is fully developed $t_*$, since they are separated by less than one Hubble time. Therefore, we generically indicate with $T_*$ the temperature at which the generation mechanism for the magnetic field takes place. 

After the end of turbulence the magnetic field simply stays frozen in the fluid at scales larger than the dissipation scale $\ti l_{\rm diss}(T_{\rm fin})\simeq \ti L(T_{\rm fin})$. Eventually other dissipation processes, due to radiation viscosity, become active \cite{dissip}.

\FIGURE[ht]{ 
	\epsfig{width=10cm, file=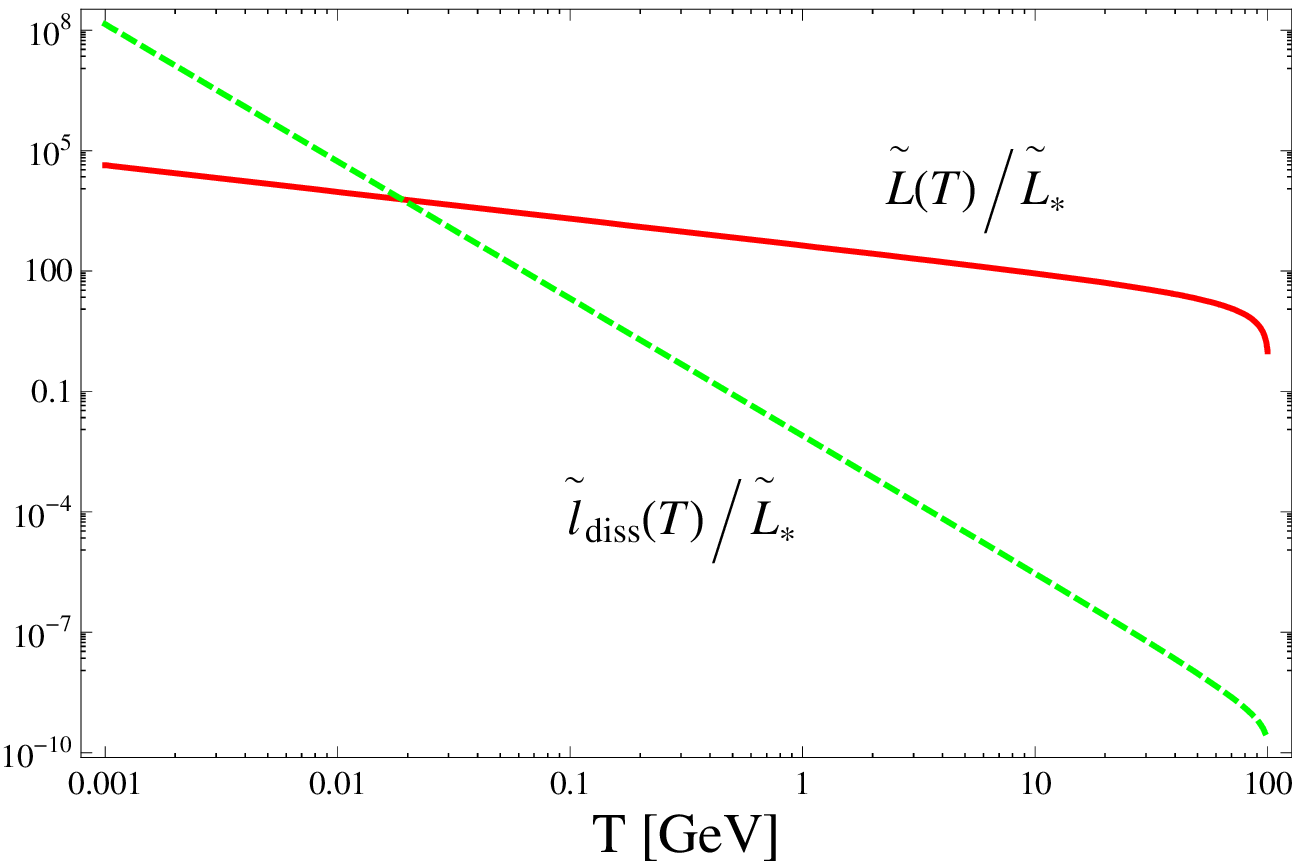}
	\caption{The evolution of the comoving magnetic correlation length (red solid line) and dissipation length (green dashed line), 
        in the inverse cascade phase for the EW phase transition. 
  Both quantities have been normalized with respect the initial value of the comoving correlation length $\ti L_*$.
        }
	\label{fig:ldiss}  
	}


\section{The GW spectrum}\label{s:spec}

\subsection{Generation of GWs from sources}

In this section we calculate the spectrum  of the GWs
generated by a helical magnetic field. We restrict our analysis to the case of maximally helical fields. 
This calculation has also been performed
in Ref.~\cite{tina}, where however stationarity in time of both the source and the GWs has been assumed.
This invariance under time translation of the source seems to us justified only if the 
time over which the source is active is much shorter than a Hubble time.
As we have argued in Section~\ref{s:setup} and derived in 
Appendix~\ref{A:Tfin}, the Reynolds number 
remains larger than unity for the scales of interest over many Hubble times.
Therefore we want to re-calculate the GW spectrum without 
the assumption of stationarity. We shall then compare our results with 
Ref.~\cite{tina}. 

The parity invariant part the GW spectrum, which is the
part which contributes to the energy density, is of the 
form~\cite{Caprini:2006jb,Caprini:2003vc}
 \bea  
  \lan  \dot h_{ij}(\bk,t)  \dot h^*_{ln}(\bq,t) \ran &=& \frac{(2\pi)^3}{4}
       \MM_{ijln}( {\bf \hat k}) \delta^3(\bk-\bq)  |\dot h|^2(k,t)   ~.
\label{hdotspec}
\eea
Here $  \MM_{ijln}({\bf \hat k})$ is the GW polarization tensor 
normalized such that $  \MM_{ijij} =4$ and 
 $|\dot h|^2(k,t)$ is related to the GW energy spectrum as 
follows~\cite{Wei,Caprini:2006jb,Caprini:2003vc}:
\bea
\ti\rho_{\rm GW}(t) &=& \frac{\lan\dot h_{ij} \dot h_{ij}\ran}
      {32\pi G a^2(t)}=
      \int_0^\infty \frac{d k}{k} 
      \frac{d\ti\rho_{\rm GW}(k,t)}{d \log k} ~,
      \nonumber \\
\frac{d\ti\rho_{\rm GW}(k,t)}{d \log k} &=& \frac{k^3 
      |\dot h|^2(k,t)}{8(2\pi)^3Ga^2(t)} ~. \label{e:rhoGW}
\eea
Our definition of the metric perturbations $h_{ij}$ differs by a 
factor $2$ with the one in~\cite{Caprini:2006jb,Caprini:2003vc} and agrees 
with~\cite{Wei}, $ds^2 = a^2(t)\left[ -dt^2 + (\de_{ij}+h_{ij})dx^idx^j\right]$.
The Fourier space expression of the projection tensor onto the transverse 
traceless component, $ \MM_{ijln}({\bf \hat k})$, is given explicitly {\it e.g.} 
in Ref.~\cite{Caprini:2003vc}.

The evolution equation which governs the generation of GWs in an expanding universe is simply, see {\it e.g.}~\cite{mybook}
\be
  \ddot h_{ij}+2\frac{\dot a}{a} \dot h_{ij}+ k^2 h_{ij} = 16\pi G a^2\Pi_{ij}  ~,
\ee
where the source term is the tensor contribution to the anisotropic stress
of the energy momentum tensor of the source. In our case these come from the magnetic 
field and  we relate the tensor anisotropic stress spectrum to the 
magnetic field spectrum in the next section.
In terms of the comoving anisotropic stress $\ti\Pi_{ij} = a^4\Pi_{ij}$ the
above equation becomes
\be
  \label{eq:h}
  \ddot h_{ij}+2\frac{\dot a}{a} \dot h_{ij}+ k^2 h_{ij} = 16\pi G 
      \frac{\ti\Pi_{ij}}{a^2}  ~.
\ee
For GWs generated by a primordial
magnetic field we cannot neglect the expansion of the Universe, since this 
source is active over a period which is much longer than a Hubble time.
In terms of the rescaled variable $\bar{h}_{ij} \equiv a h_{ij}$,
Eq.~(\ref{eq:h}) becomes
\be
  \ddot{\bar{h}}_{ij} +\lp(k^2-\frac{\ddot a}{a} \rp) \bar{h}_{ij} = 16 \pi G
      \frac{\tilde \Pi_{ij}}{a}  ~.
\ee
 In a radiation dominated background with $a\propto t$ so that 
$\ddot a =0$, 
\be
  \label{eq:hbar}
  \ddot{\bar{h}}_{ij} +k^2 \bar{h}_{ij} = 16 \pi G
     \frac{\tilde \Pi_{ij}}{a}  ~.
\ee
Since initially $h_{ij} = \dot h_{ij} = 0$, the solution of the the above 
differential equation is given by the convolution of the source 
with the retarded Green function $\GG(k,t_1,t_2) =\sin (k(t_1-t_2))$,
\be\label{e:Green}
  \bar{h}_{ij}(\bk,x<x_{\rm fin})= \frac{16\pi G}{k^2}\int_{x_\in}^{x} 
       d y ~\frac{\tilde \Pi_{ij}(\bk,y)}{a(y)} \sin (x-y) ,
\ee 
where we have introduced $x \equiv kt_1$ and $y \equiv kt_2$.

We assume that the source is active until the final time $t_{\rm fin}$ at 
which turbulence terminates and the anisotropic stress become negligible.  
This is not completely correct, since after this final time we 
have no longer an inverse cascade, but the magnetic field is frozen in and evolves according to 
flux freezing. However, this is relevant only for GW production at scales which are super-horizon at $t_{\rm fin}$, and therefore it does not affect the peak region of the GW spectrum and the value of the integrated energy density, which determines our constraints 
(see Ref.~\cite{Caprini:2006jb} and section \ref{s:spectrumhelical}).

Once the source has decayed, GWs are freely propagating. 
This behavior is described by the homogeneous solution of
Eq.~(\ref{eq:hbar}),
\be \label{eq:hhomo}
  \bar{h}_{ij}(\bk,t>t_{\rm fin})=
            A_{ij}(\bk)\sin (kt-kt_{\rm fin})+
            B_{ij}(\bk)\cos (kt-kt_{\rm fin})~.
\ee
The coefficients $A_{ij}$ and $B_{ij}$ are determined by requiring continuity 
of $\bar{h}_{ij}$ and $\dot{\bar{h}}_{ij}$ at $t=t_{\rm fin}$. 
Matching (\ref{eq:hhomo}) to the result from Eq.~(\ref{e:Green}) yields
\bea
&&  A_{ij}(\bk) = \frac{16\pi G}{k^2}\int_{x_\in}^{x_{\rm fin}}  dy ~\frac{\tilde \Pi_{ij}(\bk,y)}{a(y)} \cos (x_{\rm fin}-y)  ~, 
     \nn \\ 
&&  B_{ij}(\bk) = \frac{16\pi G}{k^2}\int_{x_\in}^{x_{\rm fin}} dy ~\frac{\tilde\Pi_{ij}(\bk,y)}{a(y)} \sin(x_{\rm fin}-y)  ~.
\eea
With Eq.~(\ref{hdotspec}), using the above solution for $\bar h_{ij}$, 
we obtain for $t>t_{\rm fin}$
\bea
&&  |\dot h|^2(k,t> t_{\rm fin}) 
      = \frac{1}{2a^2}(k^2+\HH^2)\lp(\lan A_{ij}A_{ij}^*\ran +
     \lan B_{ij}B_{ij}^*\ran \rp) 
      \nn \\ 
  &&    = \frac{(k^2+\HH^2)}{2a^2} \lp(\frac{16\pi G}{k^2}\rp)^2 
      \int_{x_\in}^{x_{\rm fin}} dy \int_{x_\in}^{x_{\rm fin}} d z  ~\cos(z-y) 
      \frac{\tilde\Pi_B(k,y,z)}{a(y) a(z)}  ~, \label{e:hPix}
\eea
where we have set $y=kt_1$ and $z=kt_2$. Furthermore, we have introduced the 
anisotropic stress unequal time power spectrum,
\be 
  \lan \tilde\Pi_{ij}(\bk,t_1) \tilde\Pi^*_{ij}(\bq,t_2)\ran = 
     (2\pi)^3 \delta^3(\bk-\bq) 
     \tilde\Pi_B(k,t_1,t_2) ~.
\ee
To obtain Eq.~(\ref{e:hPix}) we have not only performed an ensemble average, 
but also averaged over several periods so that $\lan\sin^2(kt)\ran = 
\lan\cos^2(kt)\ran = 1/2$ and $\lan\cos(kt)\sin(kt)\ran = 0$.
At times $t$ at which we can observe a GW with wave number
$k$, the latter must be largely sub-horizon so that $kt\gg 1$. We therefore may
neglect the second term in the pre-factor $(k^2+\HH^2) \simeq k^2(1+1/(kt)^2)$.
Rewriting Eq.~(\ref{e:hPix}) as integral over time, we find 
with~(\ref{e:rhoGW})
\be\label{e:rhoGWPi}
\frac{d\rho_{\rm GW}(k,t)}{d\log k} = 
      \frac{2 G }{\pi a^4(t) }k^3 
     \int_{t_\in}^{t_{\rm fin}} dt_1
     \int_{t_\in}^{t_{\rm fin}} dt_2
     ~\cos(kt_1-kt_2) 
     \frac{\tilde\Pi_B(k,t_1,t_2)}{a(t_1) a(t_2)}   ~.
\ee

\subsection{Magnetic anisotropic stresses}

According to Eq.~(\ref{e:rhoGWPi}), in order to determine the GWs produced by a cosmic magnetic 
field, we need to calculate the unequal time correlator of the 
tensor type magnetic 
stress, $\tilde\Pi_B(k,t_1,t_2)$, which  sources GWs. 
By statistical isotropy, the tensor type magnetic stress 2-point function 
has the same tensor structure as the one of GWs,
\be
  \lan \ti\Pi_{ij}(\bk,t)\ti\Pi_{ln}(\bq,t)\ran = \frac{(2\pi)^3}{4}
    \delta^3(\bk-\bq) 
    \lp[ \MM_{ijln}(\hat \bk) \ti\Pi_B(k,t,t) + \mathcal{A}_{ijln}(\hat \bk) 
\ti \Pi_A(k,t,t)\rp]
    ~.
\ee
In the above expression $\ti \Pi_A(k,t,t)$ is the term of odd parity due to the non-vanishing helicity of the magnetic field. It 
it does not  contribute to the GW energy 
density but only to their polarization ~\cite{Caprini:2003vc}. The odd parity 
projection tensor is also given in~\cite{Caprini:2003vc}.\\
Following \cite{Caprini:2003vc}, we use Wick's theorem to reduce this
four point correlator to the convolution of two 2-point correlators.
The ansatz~(\ref{e:Bansatz}) then gives for the equal time correlator
\be
  \ti\Pi_B(k,t,t) = \NN_1 \int {\rm d}^3 q 
  	\lp[ (1+\gamma^2)(1+\alpha^2) S(q,t) S(|\bk-\bq|,t) 
   +4 \gamma \alpha A(q,t) A(|\bk-\bq|,t) \rp]\label{PiGeneral}
\ee
where we set $\NN_1= 2/(4\pi)^5$, $\alpha\equiv \hat{\bf k}\cdot
 (\widehat{\bf k-q})$ and $\gamma\equiv \hat{\bf k}\cdot \hat{\bf q}$.
 In the case of a maximally helical magnetic field, 
the symmetric and antisymmetric parts of the magnetic field spectrum 
are equal on sub-horizon scales,
\be
   | A(k,t) | = S(k,t)   ~, \quad kt> 1 \,.
\ee
On super-horizon scales helicity is suppressed (see {\it e.g.}~\cite{causal}).
In order to account for this dependence, we introduce the function 
$\Sigma(t,q,|\bk-\bq|)$ in the integral~(\ref{PiGeneral})
\be
	\Sigma(t,q,|\bk-\bq|) =
	 \left\{ \begin{array}{ll} 
     1~, & \qquad \mbox{for } qt \ge 1 \mbox{ and }  |\bk-\bq| t \ge 1\,, \\    
		0~,  & 	\qquad \mbox{otherwise} \, ,
      \end{array} \right.  
\ee
and we set
\be   \label{e:Pitt}
  \ti\Pi_B(k,t,t) \simeq \NN_1 \int {\rm d}^3 q \lp[ (1+\gamma^2)(1+\alpha^2)
+4 \gamma \alpha\Sigma(t,q,|\bk-\bq|)\rp]  S(q,t)  S(|\bk-\bq|,t)\,.
\ee
The integral (\ref{e:Pitt}) for the equal time correlator is evaluated
numerically and the results is approximated by an analytical fit. 
More details on this are given in Appendix~\ref{A:Pi}. Here we simply present
the results for
two exemplary values of the spectral index, a causal spectrum with $n=2$ and a red 
spectrum with $n=-1.8$:
\bea
  \label{pi4K}
 \ti\Pi_B(K,t)&\simeq& \frac{\NN_2}{2\pi} \ti L^3(t) \ti\rho_B^2(t)  
   \frac{0.034}{1+\lp( K/12\rp)^4+\lp( K/6\rp)^{7/2}  } \,,
  \qquad n=2  ~,\\
  \label{piRed}
  \ti\Pi_B(K,t)&\simeq& \frac{\NN_2}{2\pi} \ti L^3(t) \ti\rho_B^2(t) 
\frac{(K/40)^{-3/5}}{1+(K/1.4)^{29/10}} \,, \qquad n=-1.8  ~. 	\label{pi0.2K}
\eea 
Here
\[
 \NN_2 \equiv  \lp[\frac{\Gamma\lp( \frac{2n+7}{4} \rp)}
{\Gamma\lp( \frac{1}{4} \rp)\Gamma\lp( \frac{n+3}{2} \rp)} \rp]^2 =\left\{
\begin{array}{ll}
 0.11 & \mbox{ for } n=2~, \\
 0.04 & \mbox{ for } n=-1.8 ~. \end{array} \right.
 \]

\subsection{The GW spectrum produced by a maximally helical 
magnetic field} \label{s:spectrumhelical}

Let us now consider a magnetic field with maximal initial helicity $h_B=1$,
which immediately ({\it i.e.}~at $t_*=t_{\rm in}+t_L^*$) develops an inverse cascade. 
To compute the GWs produced by this field, we have
to make assumptions about the unequal time correlator of the 
anisotropic stress. There are different possibilities which are discussed 
in the literature~\cite{CRTG,bubbles}.

We consider a completely coherent source, namely a source
with deterministic time evolution for which the unequal time correlator
is just the product of the square root of the equal time correlators at 
the different times,
\be
  \ti\Pi_B(k,t_1,t_2)  = \sqrt{\ti \Pi_B(k,t_1,t_1) }
       \sqrt{\ti \Pi_B(k,t_2,t_2) }  ~.
\ee 
This is not only the simplest approximation, but the results obtained in 
this case are also quite close to the results from a model with exponential
decoherence as discussed in Ref.~\cite{CDS2}. Furthermore, for colliding 
bubbles where numerical simulations exist, the totally coherent 
approximation is in good agreement with the numerical results~\cite{CRTG}.
This justifies our hope that this approximation captures the main features of
the resulting spectrum and, especially, that it gives a good estimate for 
the total GW energy density. Note also that this assumption has usually been made in previous works, for a magnetic field which is simply redshifting with the expansion of the universe \cite{Caprini:2001nb,Kosowsky:2001xp}. A comparision of different 
approximations can be found in Refs.~\cite{CRTG,bubbles}.

For a completely coherent source we obtain
\begin{eqnarray}
\label{drhoGW}
\frac{d\rho_{\rm GW}(k,t)}{d\log k} &\simeq& 
      \frac{2 G }{\pi a^4(t) }k^3 
     \lp\{ \lp[\int_{t_\in}^{t_{\rm fin}} dt'
     ~\cos(kt')\frac{\sqrt{ \ti \Pi_B(k,t')}}{a(t')} \rp]^2 +   
     \nn   \rp. \\  &&  \hspace*{2cm}  \lp.
     \lp[\int_{t_\in}^{t_{\rm fin}} dt' 
     ~\sin(kt')\frac{\sqrt{ \ti \Pi_B(k,t')}}{a(t')} \rp]^2 \rp\}  ~.
\end{eqnarray} 

In order to compute the above integrals, we substitute approximations
(\ref{pi4K}) for the anisotropic stresses of a magnetic field with a blue 
spectrum ($n=2$), or (\ref{pi0.2K}) for a red spectrum ($n=-1.8$).

We fix the final time at which the source of GWs ceases to be active
as the time given by the end of turbulence, when 
$\Re(L(T_{\rm fin}^{(1)}))\simeq 1$. This corresponds to the time at which 
the inertial range  ($K\gtrsim 1$) is entirely dissipated, when the 
dissipation scale has grown to reach the correlation length,
$\ti l_{\rm diss}(T_{\rm fin}^{(1)}) \simeq \ti L(T_{\rm fin}^{(1)})$. This condition 
determines the value of 
the final temperature $T_{\rm fin}^{(1)}$ at which turbulence terminates 
(in principle, the magnetic field is not damped after this temperature, but simply stays frozen in the fluid and keeps on generating GWs; however, here for simplicity we restrict to GW production during the turbulent phase, an assumption which, as previously mentioned, does not affect our result in a relevant way). 

In Appendix~\ref{A:Tfin} we estimate the final temperature for 
inverse cascade turbulence initiated at different times:
\bea
T_{\rm fin}^{(1)} &\simeq& 21\,{\rm MeV} \mbox{ for the EW phase transition,} \label{TFINEW} \\ 
T_{\rm fin}^{(1)} &\simeq& 5\,{\rm MeV} \mbox{ for the QCD phase transition, }\label{TFINQCD} \\ 
T_{\rm fin}^{(1)} &\simeq& 1\, {\rm GeV} \mbox{ for inflation.} \label{TFININ}
\eea
The final time $t_{\rm fin}^{(1)}$ corresponding to these temperatures is 
given by~\cite{mybook}
\be
   t_{\rm fin}^{(1)} = t(T_{\rm fin}^{(1)}) \simeq 0.5\,
   \lp(g_{\rm eff}(T_{\rm fin}^{(1)})\rp)^{-1/6}\,
   \frac{m_{Pl}}{T_0T_{\rm fin}^{(1)}} \,.
\ee
On the other hand, we know that the dissipation scale grows more rapidly 
than the correlation scale. Therefore, when a given wavelength, smaller than the 
 correlation scale but initially larger than the dissipation scale,
becomes of the order of the dissipation scale 
$2\pi/k \simeq \ti l_{\rm diss}(T_{\rm fin}^{(2)}(k))$, turbulence is 
dissipated on this scale and the GW source has decayed. 
This defines a second, $k$-dependent 
final temperature  $T_{\rm fin}^{(2)}(k) \ge T_{\rm fin}^{(1)}$ given by
$$ \ti l_{\rm diss}(T_{\rm fin}^{(2)}(k)) =2\pi/k=\ti L_*/K_* \,. $$
Since $T_{\rm fin}^{(2)}(k)> T_{\rm fin}^{(1)}$, the final time of integration for 
the wave number $k$ is $t_{\rm fin}^{(2)}(k)=t(T_{\rm fin}^{(2)}(k)) <
t_{\rm fin}^{(1)}$.

In Appendix~\ref{A:ldiss} we derive analytical expressions for 
$t_{\rm fin}^{(2)}(k)$, taking into account the time evolution of the 
dissipation length, see Eqs.~(\ref{e:tk2EW}) and (\ref{e:tk2inf}).
The final time $t_{\rm fin}$ is given by
\be\label{e:tfin}
   t_{\rm fin}(k) = {\rm min} \lp[ t_{\rm fin}^{(1)}, t_{\rm fin}^{(2)}(k)\rp]  ~.
\ee
Indeed, for scales smaller than $\ti L(t_{\rm fin}^{(1)})$, hence 
$K(t_{\rm fin}^{(1)})>1$, $t_{\rm fin}(k)$ is equal to $t_{\rm fin}^{(2)}(k)$, while 
larger scales are dissipated only at the end of turbulence, $t_{\rm fin}^{(1)}$.

The $K_*$-dependence of $t_{\rm fin}$  for magnetic fields generated at the EW phase transition and at 
inflation is plotted in Fig.~\ref{fig:tfin}.  The 
final time starts to decrease  for small wavelengths, namely  around 
$K_* =K_1^*\simeq 10^{-3}$ for the EW transition and at 
$K_*=K_1^* \simeq 10^{-9}$ for inflation. This value is given by 
\be
1= K_1(T_{\rm fin})=K_1^* \frac{\ti L(T_{\rm fin})}{\ti L_*} = K_1^*\tau_{\rm fin}^{2/3} 
\simeq K_1^*\left(\frac{2v_L}{\ep}\frac{T_*}{T_{\rm fin}}\right)^{2/3}\,,
\ee
so that
\be 
K_1^* \simeq \left(\frac{\ep}{2v_L}\frac{T_{\rm fin}}{T_*}\right)^{2/3}\,.
\ee

\FIGURE[ht]{ 
\epsfig{width=10cm, file=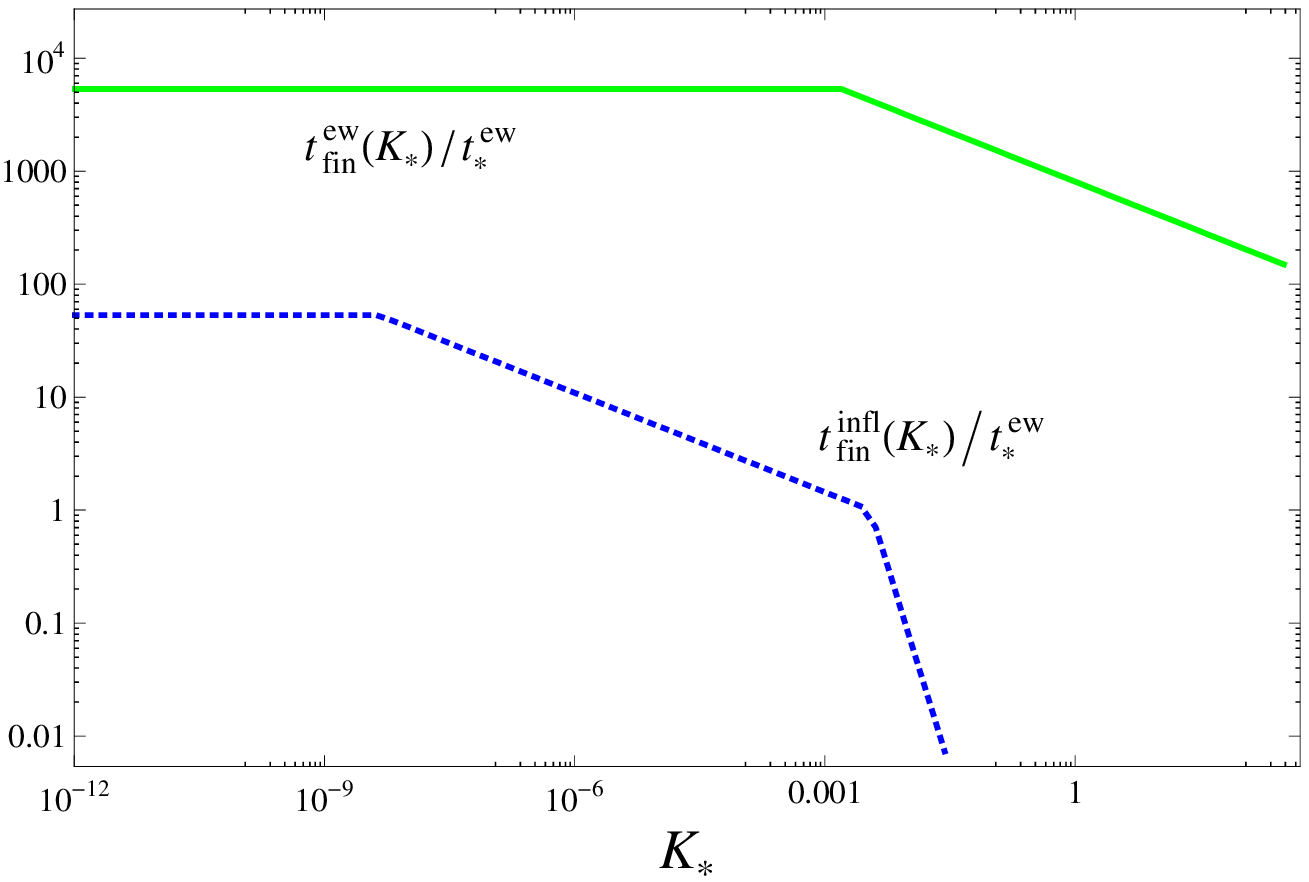}\\
\caption{$K_*$-behavior of $t_{\rm fin}$ 
 for the EW phase transition (green, solid line) and 
for inflation (blue, dotted line), both normalized with respect $t_*^{\rm ew}$.}
\label{fig:tfin}       
} 

\vspace{1cm}

To calculate the integrals~(\ref{drhoGW}) we use Eqs.~(\ref{pi4K}) 
and (\ref{pi0.2K}) for the time evolution of the magnetic anisotropic stress. More details on the explicit form of (\ref{drhoGW}) are given in 
Appendix~\ref{A:gwint}. The GW energy spectrum can be written in the form
\be
  \frac{d\Omega_{\rm GW}(k,t_0)}{ d\log k} \simeq 3\,
		 \NN_2 \lp( \frac{\bar g}{g_0}\rp)^{1/3}
			 \frac{(\ti\Omega_B^*)^2}{\Omega_{\rm rad}} \, \II_{\rm GW}(K_*)  ~,
\ee
which holds for magnetic fields with both a blue and a red spectrum. Here $\bar g$ denotes an average number of relativistic degrees of freedom while the source is active. We neglect this factor in the following (it enters the limits for the magnetic field amplitude only as $(\bar g/g_0)^{1/12}$). 
The integral $ \II_{\rm GW}(K_*)$ determines the spectral shape. We have calculated it numerically for both cases: a magnetic field generated at the EW phase transition with a blue
spectrum, with parameters $n=2$, $\ep=0.01$ and $v_L^2=0.2$, and one generated at inflation 
with a red spectrum, setting the parameters to $n=-1.8$ and $\ep= v_L= 1$. This means that in the inflationary case, the initial stirring scale is set equal to the horizon size, and the eddy turnover time is half the initial Hubble time, $t_L^*=t_{\rm in}/2$ \footnote{Note that in the inflationary case, we could as well have chosen the stirring scale to coincide with the horizon at any time. One could argue, in fact, that as soon as a scale enters the horizon, it causes a stirring of the cosmic fluid, acting as the stirring scale. We have evaluated the GW spectrum also setting $\ti L(\tau)\propto \mathcal{H}^{-1}\propto \tau$ (instead of $\propto \tau^{2/3}$) and consequently $\ti\rho_B(\tau)\propto \tau^{-1}$ to maintain the inverse cascade. We did not find an appreciable difference among the two resulting GW spectra.}.  

For a causal magnetic field spectrum with $n=2$, the GW 
density parameter in units of $(\ti\Omega_B^*)^2/\Omega_{\rm rad}$ is shown 
in Fig.~\ref{fig:GWK}.
\FIGURE[ht]{ 
\epsfig{width=10cm, file=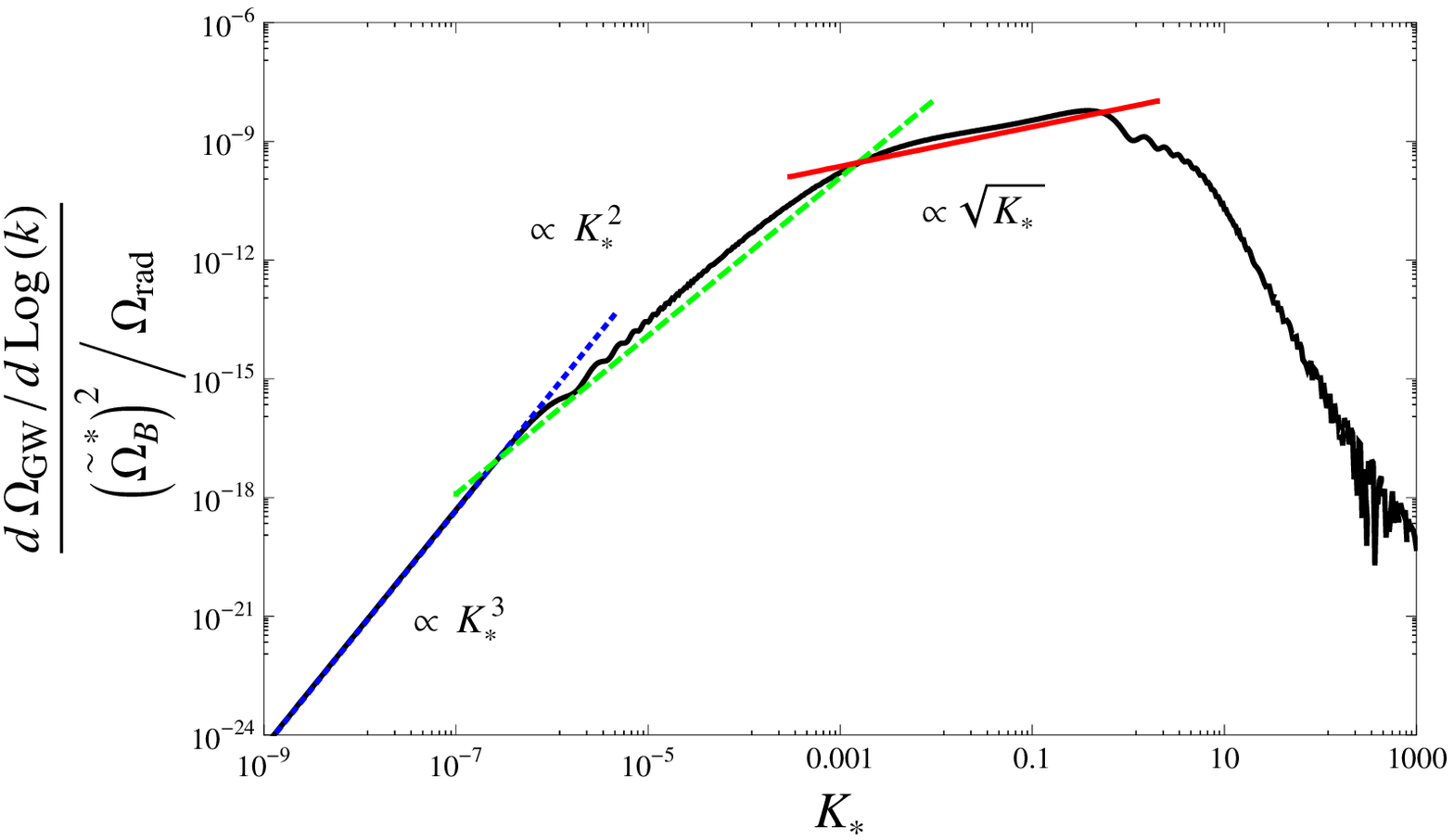}\\
\caption{The GW energy density spectrum from a causal 
magnetic field $n=2$ generated at the EW phase transition with maximal initial
helicity $h_B=1$. The spectrum grows like $k^3$ (blue, dotted line) up to 
the horizon at the end of the turbulent phase, $k\simeq t_{\rm fin}^{-1}$. Then 
the slope is given by $k^2$ (green, dashed line) up to the initial horizon 
$k\simeq t_*^{-1}$ and by $\sqrt{k}$ (red, solid line) up to the peak at 
$k\simeq (2\pi)/t_L^*$.  Above the peak frequency the GW spectrum decays 
rapidly. For the EW phase transition one has $f/{\rm mHz}=4\,K_*$.}
\label{fig:GWK}       
} 
Below the peak frequency, located at $k\simeq 2\pi/t_L^*$, the numerical result can be approximated by
\bea
\label{OmGWEW}
\frac{ d \Omega_{\rm GW}(k,t_0)}{d\log k} \simeq 
	3\, \NN_2 \frac{(\ti\Omega_B^*)^2}{\Omega_{\rm rad}}
   \left\{ \begin{array}{ll} 
\epsilon_1 K_*^3   & \quad \mbox{for } 0<K_* < \ti L_*/(2\pi\, t_{\rm fin}) \,,  \\
	 	\\       
\epsilon_2 K_*^2  & \quad \mbox{for  } \ti L_*/(2\pi\, t_{\rm fin})<K_*< \ti L_*/(2\pi\, t_*) \,,\\  
	 	\\  
	\ep_3 K_*^{1/2}  &  \quad \mbox{for }  {\ti L}_*/(2\pi\, t_*)<K_*<\ti L_*/ t_L^*  ~.
     \end{array} \right.
\eea
A part from the peak, the spectrum shows features at wave numbers corresponding to the characteristic times of the source: $k\simeq t_{\rm fin}^{-1}$ and $k\simeq t_*^{-1}$ ($t_{\rm in}$ is too close to $t_*$ to be distinguishable). More details on the fit, including the values of the parameters $\ep_i$, are given in Appendix \ref{app:fits}. Using the above approximation (\ref{OmGWEW}), we find that the integrated energy density parameter of GWs 
today is approximately given by
\bea
	\label{finalGWblue}
\Omega_{\rm GW}(t_0) = \int_0^{\infty} \frac{d k}{k} \frac{ d
   \Omega_{\rm GW}(k,t_0)}{d\log k}
		 \simeq 6\, \NN_2 \frac{(\ti\Omega_B^*)^2}{\Omega_{\rm rad}} 
		 \, \ep_3 \sqrt{\frac{\ti L_*}{t_L^*}}  \simeq 2 \times 10^{-8} \,\frac{(\ti\Omega_B^*)^2}{\Omega_{\rm rad}}~.
\eea
Here we have neglected the decaying part of the GW spectrum,
since the main contribution to the integrated energy density $\Omega_{\rm GW}$ comes from the part of the spectrum
 close to the peak $k \simeq 2\pi/t_L^*$. For the numerical value, we have inserted $\ep=0.01$ and $v_L^2=0.2$ for a magnetic field generated at the EW phase transition with $T_*=100\,$GeV. The GW spectrum from a magnetic field generated at the QCD phase transition, $T_*=100$ MeV, for the same values of $\ep$ and $v_L$ is very similar, and in particular it has the same amplitude at the peak. 

Next we consider an inflationary magnetic field with red spectrum $n=-1.8$. In this case 
we have to use the anisotropic stress given in Eq.~(\ref{pi0.2K}). 
The resulting spectrum is plotted in Fig.~\ref{fig:redGWK}, and below the peak frequency it can be approximated by 
\bea \label{OmGWINF}
\frac{{\rm d}\Omega_{\rm GW}(k,t_0)}{{\rm d}\log k} \simeq 
	3\,\NN_2 \frac{(\ti\Omega_B^*)^2}{\Omega_{\rm rad}}
      \left\{ \begin{array}{ll} 
	\epsilon_4 K_*^{2n+6} ,   & ~ \mbox{for } 0<K_*< \frac{ \ti L_*}{2\pi \,t_{\rm fin}}  \,,  \\
	 	\\       
	\ep_5 K_*^{(2n+10)/3}  ,  & ~ \mbox{for  } \frac{ \ti L_*}{2\pi \,t_{\rm fin}} < K_* < \lp(\frac{(4\pi)^5}{R_*^9}\rp)^{1/7} , \\
     		\\
	\ep_6 K_*^{-(2+6n)/5}  ,  &  ~\mbox{for  } \lp(\frac{(4\pi)^5}{R_*^9}\rp)^{1/7} < K_* < \frac{\ti L_*}{t_L^*} \simeq 2  ~.
     \end{array} \right.
\eea
Here, the $K_*-$dependence is written in terms of the general spectral index $n$, and it is valid for any $n<-3/2$. On the other hand, the values of the matching constants $\ep_i$ are derived in Appendix \ref{app:fits} under the assumption $n=-1.8$. 

\FIGURE[ht]{ 
\epsfig{width=9.5cm, file=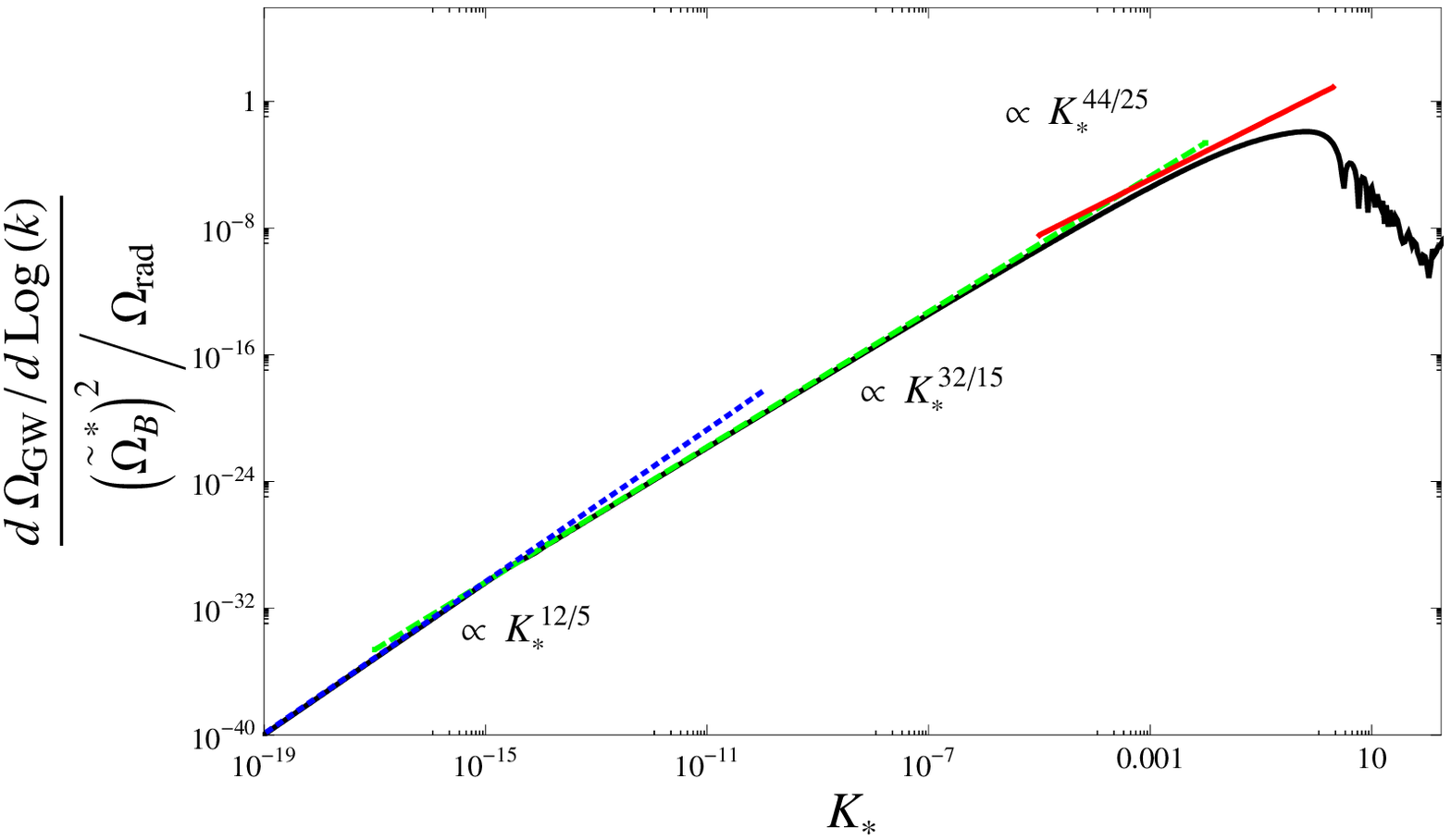}\\
\caption{GW energy density spectrum from an inflationary magnetic field with red spectrum $n=-1.8$ and 
maximal initial helicity $h_B=1$. The
analytically expected behavior is also indicated (\cf~Eq.~(\ref{OmGWINF})):
$\propto K_*^{2n+6}$ (blue, dotted line), $\propto K_*^{(2n+10)/3}$ (green, 
dashed line) and $\propto K_*^{-(2+6n)/5}$ (red, solid line). For inflation one has $f/{\rm Hz}\simeq 10^7\,K_*$.}
\label{fig:redGWK}       
} 

Neglecting the decaying part of the GW spectrum, and using the slope $K_*^{-(2+6n)/5}$ up to the peak $k\simeq 2\pi/t_L^*$, we find the total GW energy density parameter:
\bea
	\label{finalGWred}
\hspace*{-0.6cm}	\Omega_{\rm GW}(t_0)  &\simeq& 3\,  \NN_2 
	\frac{(\ti\Omega_B^*)^2}{\Omega_{\rm rad}}
	\lp(\frac{-5}{2+6n}\rp) \ep_6  
        		 \lp(\frac{\ti L_*}{t_L^*} \rp)^{-(2+6n)/5} \quad \quad   n<-\frac{3}{2} \\
& \simeq& 5.2 \, \frac{(\ti\Omega_B^*)^2}{\Omega_{\rm rad}}\,,\nn 
 \eea
Note that this approximation causes an overestimation of the total GW energy density of about three orders of magnitude. However, this does not affect the bounds on the magnetic field amplitude significantly: 
it translates into a bound that is stronger by about the 20\% 
(see~section~\ref{s:res}). For the numerical value in the last equality of (\ref{finalGWred}), we have inserted the value $n=-1.8$ and $\ep=v_L=1$. 

\vspace*{0.6cm}

Finally, to make contact with future observations, we express the 
GW spectra in terms of 
the GW amplitude as function of the frequency $f$.
For this we use~\cite{michele}
\be
  h(f)=1.26 \times 10^{-18} h_0 \sqrt{\Omega_{\rm GW} (f)}\lp(\frac{\rm Hz}{f} \rp)~,
\ee
where the frequency is $f=k/(2\pi)$ and 
$$
\Omega_{\rm GW} (f) \equiv \left.\frac{{\rm d}\Omega_{\rm GW}(k,t_0)}{{\rm d}\log k}
  \right|_{k=2\pi f} \,.
$$
The behavior of $h(f)$ for a causal and for an inflationary produced magnetic field, 
choosing a maximal magnetic field amplitude of 
$\Omega_{B}^* \simeq 0.1$, is plotted in 
Fig.~\ref{f:freqK}. For another  magnetic field  density parameter the resulting amplitude $h(f)$
is simply rescaled by the factor
$\Omega_{B}^*/0.1$.

\FIGURE[ht]{ 
\epsfig{width=10cm, file=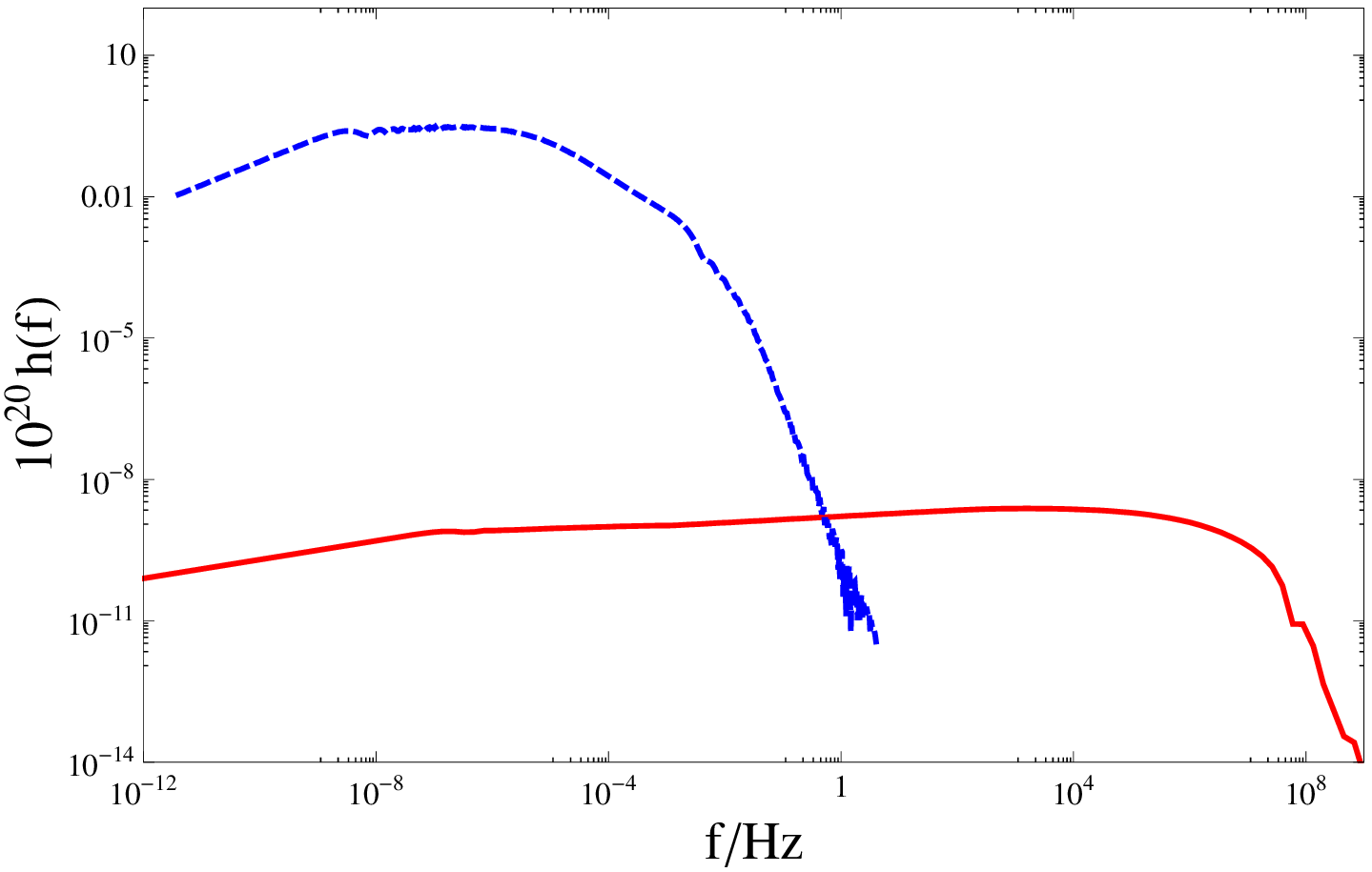}\\
\caption{The GW amplitude $h(f)$ as a function of frequency from a maximally helical magnetic field with $\Omega_{B}^* \simeq 0.1$. Blue, dotted: causal generation at the EW phase transition with $n=2$; red, solid: generation during inflation with $n=-1.8$. }
\label{f:freqK}       
} 

The result for a magnetic field with a blue spectrum, shown in Fig.~\ref{f:freqK}, agrees qualitatively with the one obtained in Ref.~\cite{tina} . The position of the peak is the same, although the amplitude is higher in Ref.~\cite{tina} by nearly one order of magnitude. This may come from the different assumptions for the unequal time correlator: in \cite{tina} the source is assumed to be stationary with exponential decorrelation, while we choose a completely coherent source. Moreover, in the magnetic energy spectrum, we have modelled the transition from the $k^{n+2}-$behaviour at low wave number to the inertial range, while Ref.~\cite{tina} extends the Kolmogorov spectrum down to $k=2\pi/\ti L_*$. We have experienced that this leads to a significant overestimation of the peak amplitude \cite{Caprini:2006jb}.
Furthermore, the low frequency tail of the spectrum in \cite{tina} grows as $\sqrt{f}$, whereas in our case it becomes constant for $f>1/t_{\rm fin}$: this is due to the fact that our source is long lasting, while theirs lasts for less than one Hubble time (see \cite{CDS2}).

For $\Omega_{B}^* \simeq 0.1$, the EW result is somewhat below the sensitivity range of LISA \cite{LISA}. On the other hand, the inflationary result is much below any proposed experiment: even though the energy density is higher, much of it is at high frequency, resulting in a very low gravitational wave amplitude (see Fig.~\ref{f:freqK}).  

\subsection{The peak position\label{s:peak}}

In Ref~\cite{CRTG} it has been argued that approximating the source of GWs by a 
discontinuous function might influence the decay law and even the peak 
position of the resulting GW spectrum. Therefore, as already discussed in section \ref{s:dirandinv}, 
in this work we model the 'switching on' process and avoid
a discontinuous source function. Here we compare our result with what we would have obtained assuming a discontinuous source.

Fig.~\ref{fig:peak} shows the results for a continuous source and for a 
discontinuous one where the inverse cascade starts instantaneously 
at $ t = t_*$. The predictions of Ref.~\cite{CRTG} are 
confirmed: in the discontinuous case the peak is no longer at $t_L^*$ but
rather at $\ti L_* \simeq t_L^*/(2v_L)$, leading to an over estimate of the resulting
GW output. Therefore, it is important to take into account continuity.

In order to further clarify this issue, we vary the initial speed $v_L$, which relates the initial correlation length 
$\ti L_*$ and the characteristic turnover time $t_L^*\simeq \ti L_*/(2v_L)$. We fix it to $v_L = 10^{-2}$ so that $\ti L_*$ and $t_L^*$ are clearly separated: $t_L^*\simeq 10^{2} \ti L_*$. As can be seen in Fig.~\ref{fig:peak}, in the discontinuous case the peak position is independent of the velocity $v_L$, while in the continuous one it is located at $k=2\pi/t_L^*$. 

Summarising, having assumed that the magnetic field processed by MHD turbulence needs a characteristic time of order
$t_L^*$ to 'form' induces a peak in the GW spectrum at a wave number corresponding to this characteristic time.  
If the field builds up much faster, almost instantaneously, the peak can move to $k=2\pi/\ti L^*$.

\FIGURE[ht]{ 
\epsfig{width=9.5cm, file=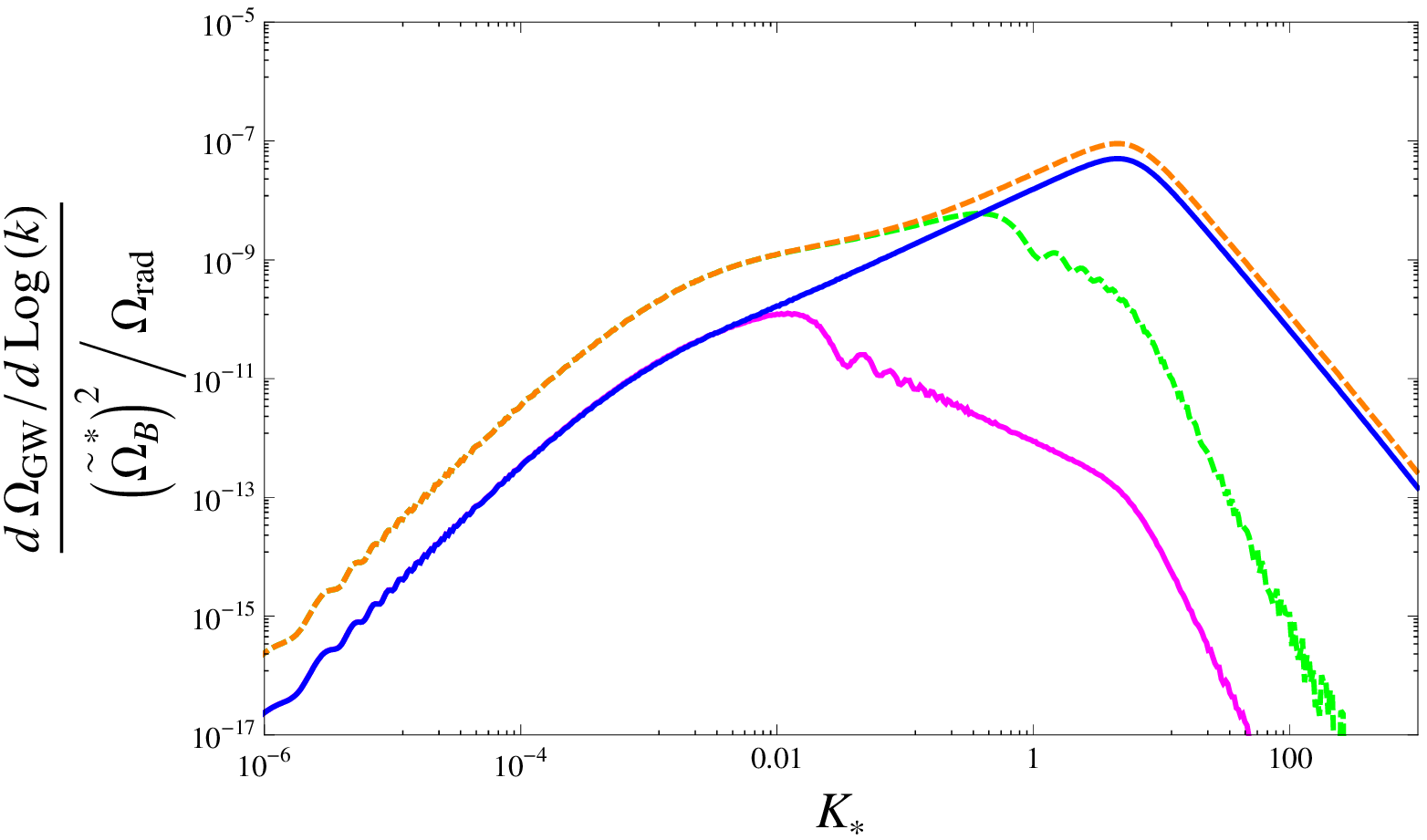}\\
\caption{The GW spectrum, normalised to the magnetic energy density, for a causal 
magnetic field $n=2$ with maximal initial helicity $h_B=1$. Continuous source: green, dashed line with $v_L^2=0.2$ and magenta, solid line with $v_L^2=10^{-4}$. Discontinuous source: orange, dashed line with $v_L^2=0.2$ and blue, solid line with $v_L^2=10^{-4}$. In the continuous case the spectrum peaks at $k \simeq 2\pi/t_L^*$ while in the discontinuous one it peaks at $k \simeq 2\pi/\ti L_*$.}
\label{fig:peak}       
} 

\section{Limits}\label{s:res}

The main aim of this paper is to derive constraints on the amplitude of a primordial 
magnetic field scaled to today, smoothed over an interesting characteristic scale. 
We choose the comoving scale  $\ti\lambda \simeq 0.1$ Mpc. On one hand, this is approximately the 
largest correlation scale of cluster magnetic fields today, on the other hand it is only 
a little larger than the smallest scale which survived dissipation prior to 
recombination \cite{dissip}. Note however that, accounting for the full evolution of the magnetic field, Ref.~\cite{Banerjee:2004df} found a smaller dissipation scale at recombination, of about 1 kpc. 

To constrain the magnetic field amplitude, we use the GW energy density generated by the magnetic field and we apply the nucleosynthesis bound to these GWs, the same strategy followed in Ref.~\cite{Caprini:2001nb}.
We define the comoving magnetic field smoothed on the comoving scale 
$\ti\lambda$ by
\be
  \ti B_i(\ti \la)= \frac{1}{V_1} \int d^3 x \, \ti  B_i(\bx) 
       \exp \lp( -\frac{x^2}{\ti\lambda^2} \rp),
\ee  
where $V_1$ is the normalization volume given by 
\be
  V_1= \int d^3 y  \exp \lp( -\frac{y^2}{\ti\lambda^2} \rp) = 
(\sqrt{\pi}\ti\lambda)^3,
\ee 
and $\ti B$ is the magnetic field scaled to today, 
$ B(\bx,t)= \ti B(\bx)/a^2(t)$. 
A short calculation gives the smoothed amplitude
\be
  \ti B^2_\la= \frac{1}{V_2} \int d^3 x \lan\ti{\bf B}(\bx)\cd
   \ti{\bf B}(\bx+\by)\ran 
       \exp \lp( -\frac{y^2}{2\ti\lambda^2} \rp),
\ee  
whit $V_2= (\sqrt{2\pi}\ti\lambda)^3$. 
Translating the above expression to Fourier space we obtain 
\be
	\label{Blim1}
  \ti B_{\lambda}^2= \frac{1}{2\pi^2} \int_0^{\infty} dk\, k^2 S(k,t) 
        \exp \lp( -\frac{k^2\ti\lambda^2}{2} \rp)=
         8\pi\int_0^{\infty} dk\, \varepsilon_B(k,t) 
           \exp \lp( -\frac{k^2\ti\lambda^2}{2} \rp) ~.
\ee 
We relate this amplitude to the comoving magnetic density parameter $\ti\Omega_B^*$ 
using Eqs.~(\ref{epsK}),~(\ref{normalizationK}) for $\varepsilon_B(k,t)$ which are valid  
for both direct and inverse cascade. Performing the above integral 
we find
\be
  \ti B_{\lambda}^2(t)= 8\pi\rho_c(t_0) \ti\Omega_B(t)  
  	\frac{\Gamma\lp(\frac{2n+7}{4}\rp)}{\Gamma\lp(\frac{1}{4}\rp)}~
 \mathcal{U}\lp[\frac{n+3}{2},\frac{3}{4},2\pi^2
   \lp(\frac{\ti\lambda}{\ti L}\rp)^2\rp]  ~,
\ee  
where $\mathcal{U}$ denotes the confluent hyper-geometric function, see 
{\it e.g.}~\cite{AS}.

We assume that the initial helicity of the cosmic magnetic field responsible 
for GW production is maximal, $h_B=1$. Therefore, during the inverse 
cascade phase, the magnetic correlation length evolves as given in 
Eq.~(\ref{Linverse}), and the product 
$\ti L(t) \ti\Omega_B(t)$ is constant in time.
The inverse cascade goes on until the temperature $T_{\rm fin}$ at which MHD 
turbulence terminates. In the maximally helical case, this can vary from a 
temperature of a few MeV to approximately 1 GeV depending on the epoch of 
generation of the field (see Eqs. (\ref{TFINEW})-(\ref{TFININ})). Up to this final time, the correlation length has 
grown substantially, but one readily confirms that it remains several 
orders of magnitude smaller than our scale of
interest, $\ti\lambda=0.1$ Mpc $\simeq 10^{13}$sec,
\be
\ti L(t_{\rm fin})\simeq \ti L_*\left( \frac{t_{\rm fin}}{t_L^*}\right)^{2/3}
   \ll \ti\lambda  ~.
\ee
Therefore, we can expand the function 
$\mathcal{U}(a,b,z) =z^{-a}[1 +{\cal O}(z^{-1})]$. With this we find
\be
\label{Bpast}
  \ti B_{\lambda}^2(t)= \frac{8\pi\Ga\lp(\frac{2n+7}{4}\rp)}{(\sqrt{2}\pi)^{n+3}
  \Ga\lp(\frac{1}{4}\rp)} \rho_c(t_0) \ti\Omega_B(t)  
		\lp(\frac{\ti L(t)}{\ti\lambda}\rp)^{n+3}  ~.
\ee  
Taking into account that during inverse cascade $\ti\Om_B(t)\ti L(t) =$ 
constant and $\ti L(t) =\ti L_*\tau^{2/3}$,  we obtain at 
$t_*\ll t\leq  t_{\rm fin}$
\bea \label{e:Bt}
  \ti B_{\lambda}^2(t)&=& \ti B_{\lambda}^2(t_*)  \tau^{2(n+2)/3}   ~, \mbox{ with}
	\\   \label{e:Bt*}
  \ti B_{\lambda}^2(t_*) &\equiv& 
  \frac{8\pi\Ga\lp(\frac{2n+7}{4}\rp)}{(\sqrt{2}\pi)^{n+3}
  \Ga\lp(\frac{1}{4}\rp)} \rho_c(t_0) \ti\Omega_B^*  
		\lp(\frac{\ti L_*}{\ti\lambda}\rp)^{n+3}  \,.  
\eea

For $t>t_{\rm fin}$ the primordial fluid enters in the viscous phase and the 
magnetic field energy density is dissipated by radiation viscosity \cite{dissip}. During 
this phase, the evolution on large scales is similar to MHD selective 
decay, \emph{i.e.} such that the large scale part of the power spectrum 
remains constant: $\ti\Omega_B(t) \ti L^{n+3}(t)={\rm constant}$.  
From the general evolution of $\ti B_\la$ given in Eq.~(\ref{Bpast}), we  
see that it is justified to assume that, on the scale $\ti\la$, after 
$t_{\rm fin}$ the magnetic field energy density evolves only by redshifting. 
Therefore, the comoving quantity $\ti B_\lambda(t)$ remains constant:
\be
\ti B^2_\la \equiv \ti B_\lambda^2(t\geq t_{\rm fin})\simeq 
     \ti B_{\lambda}^2(t_*)\left(\frac{t_{\rm fin}}{t_L^*}\right)^{2(n+2)/3}\,.
\label{Btoday}
\ee 

Like every contribution to radiation energy density 
prior to nucleosynthesis,  $\Omega_{\rm GW}$ is constrained by the 
nucleosynthesis bound~\cite{FS}
\be
   \Omega_{\rm GW} \leq \Omega_{\rm lim}  \simeq 0.1\,\Om_{\rm rad}\,.
\ee
Via Eqs.~(\ref{finalGWblue}),~(\ref{finalGWred}) this yields a constraint on 
the magnetic field energy density parameter $\ti\Omega_B^*$, in terms of 
$\ti L_*/t_{\rm in} =\ep$ and $\ti L_*/t_L^*\simeq 2v_L$:
\bea
&\ti\Om_B^* \lesssim \frac{36}{(\ep^{11} \, v_L^2)^{1/12}}  \Om_{\rm rad} \hspace*{0.8cm}
 & \mbox{ blue case } n=2, \mbox{ EW} \,,\label{OmBEWlim} \\
&\ti\Om_B^* \lesssim 0.14\,v_L^{(2+6n)/10}\,\Om_{\rm rad}
& \mbox{ red case } n=-1.8, \mbox{ inflation} \label{OmBINFlim} \,.
\eea
With the help of Eqs.~(\ref{e:Bt})-(\ref{Btoday}) we
translate this into a constraint on $\ti B_\lambda$. Using 
$\rho_{\rm rad}(t_0) =\rho_c(t_0)\,\Om_{\rm rad} \simeq 2\times 10^{-51}$GeV$^4$
$\simeq 0.4\times 10^{-12}\,($Gauss$)^2$, we find for a blue magnetic field generated at the EW phase transition
\bea
\frac{\ti  B_{\lambda}}{1\mu{\rm Gauss}} &\lesssim& 
	\frac{0.3}{(\ep^{11} \, v_L^2)^{1/24}}   
	\lp(\frac{\ti L_*}{\ti\lambda}\right)^{(n+3)/2}
	\left(\frac{t_{\rm fin}}{t_L^*}\right)^{(n+2)/3}\, , \label{BlaEWPT}
	 \\ &&
\mbox{ for } n= 2, \mbox{ generated at the EW phase transition}\nn
\eea
and for a red magnetic field generated during inflation
\bea
\frac{\ti  B_{\lambda}}{1\mu{\rm Gauss}} &\lesssim& 
	0.27\,v_L^{(2+6n)/20}	\lp(\frac{\ti L_*}{\ti \lambda}\right)^{(n+3)/2}\!\!
 	\left(\frac{t_{\rm fin}}{t_L^*}\right)^{(n+2)/3} \, ,
 	 \\ &&
 	\mbox{ for } n=   -1.8, \mbox{ generated at inflation.}\nn
\eea
In these equations the pre-factors are calculated using the spectral indexes 
and the initial and final temperatures corresponding to the generation times. 
The dependence on $v_L$ and $\ep$ is kept explicit for completeness.

It is interesting to see that the inverse cascade simply relaxes the limit by 
the factor $\tau_{\rm fin}^{(n+2)/3}$, absent in the non-helical case for which 
$\ti B_\la(t)=\ti B_\la^*$. 
This factor tends to $1$ for 
$n\ra -2$, the limiting value for which the inverse cascade relations
(\ref{Einverse}) and (\ref{Linverse}) apply. For causal generation with $n=2$
the limit for a magnetic field is substantially reduced, while for a red 
magnetic field spectrum with $n=-1.8$ the reduction is only by 
$\tau_{\rm fin}^{0.2/3}$. 

Let us apply our findings to the two 
generation mechanisms considered above (EW phase transition and inflation), 
to which we add also the interesting case of the QCD phase transition 
\cite{QCD}. As mentioned in Sec.~\ref{s:spectrumhelical}, we have 
evaluated the spectrum also in this case, finding a very similar amplitude 
to the EW phase transition case, for the same values of $v_L$ and $\ep$. 
Therefore, we are confident that we can trivially extend the above 
Eq.~(\ref{BlaEWPT}) also to this case. 

For a maximally helical field we 
find  the following limits on $\ti B_\lambda$:
\begin{itemize}
\item
If the field is generated at the  EW phase transition at 100 GeV, 
$t_{\rm in}\simeq 7.8 \times 10^4$ sec, assuming a causal spectrum with 
$n=2$, taking the values $\epsilon=0.01$, $v_L^2=0.2$, 
$T_{\rm fin}\simeq 21$ MeV,
setting $g_{\rm eff}(T_{\rm fin}) =43/4$ 
and using $\ti\la=0.1$ Mpc $\simeq 10^{13}$ sec, we obtain the constraint  
\be\label{e:BlimEW}
      \ti B_{0.1\,{\rm Mpc}} \lesssim    8 \times 10^{-24}\, {\rm Gauss,} ~~
\mbox{ EW phase transition, } T_* =100{\rm GeV}
 \ee

\item
If the field is generated at the QCD phase transition at 100 MeV, 
$t_{\rm in}\simeq 1.1\times 10^8$ sec, with the same parameters as before but 
 $T_{\rm fin}\simeq 5$ MeV, the constraint becomes  
\be
     \ti  B_{0.1\,{\rm Mpc}} \lesssim  2\times 10^{-19}\, {\rm Gauss,}~~
\mbox{ QCD phase transition, } T_* =100{\rm MeV}
  \ee
\item
 If the field is generated during inflation at $T_*\simeq 10^{14}$GeV,
$t_{\rm in}\simeq 7 \times 10^{-8}$sec, with an acausal red spectrum 
$n=-1.8$, choosing $\epsilon=v_L=1$, $T_{\rm fin}\simeq 1$ GeV and $g_{\rm eff}(T_{\rm fin}) =287/4$, we find the constraint
 \be
      \ti  B_{0.1\,{\rm Mpc}} \lesssim  2 \times10^{-18}\, {\rm Gauss,}~~ \mbox{ inflation, } n=-1.8 
\ee
\item
 If the field is generated during inflation but with a blue, acausal 
spectrum $n=0$, with the same values of the parameters as before we 
find the stronger constraint   
 \be\label{e:Bliminf0}
     \ti  B_{0.1\,{\rm Mpc}} \lesssim  4 \times 10^{-28}\, {\rm Gauss,} ~~  \mbox{ inflation, } n=0\,.
  \ee
\end{itemize}

The above limits are summarised in Fig.~\ref{f:limits} as a function of $n$, and in Table \ref{tab}.
These upper bounds on the amplitude of the primordial magnetic field are
less stringent than the ones obtained from a direct cascade by the factor 
$\tau_{\rm fin}^{(n+2)/3}$. Moreover, they strongly depend on the choice for the smoothing scale $\ti\la$, in particular for blue spectra. The scaling with $\ti\la$ is in fact given by ({\it c.f.}~Eq.~(\ref{Bpast}) and \cite{causal})
\be
\ti B_{\la_1}=\ti B_{\la_2}\left(\frac{\ti\la_2}{\ti\la_1}\right)^{(n+3)/2}\,.
\ee
Therefore, for a smaller smoothing scale of {\it e.g.} $\ti\la=1$ kpc \cite{Banerjee:2004df}, the above bounds are relaxed by a factor of $10^5$ in the EW and QCD generation cases, and by a factor of $10^3$ in the inflationary case with flat spectrum $n=0$. For red spectra the bound does not change much, {\it e.g.} in the inflationary case with $n=-1.8$ it is relaxed only by a factor of about $16$. 

\FIGURE[ht]{ 
\epsfig{width=10cm, file=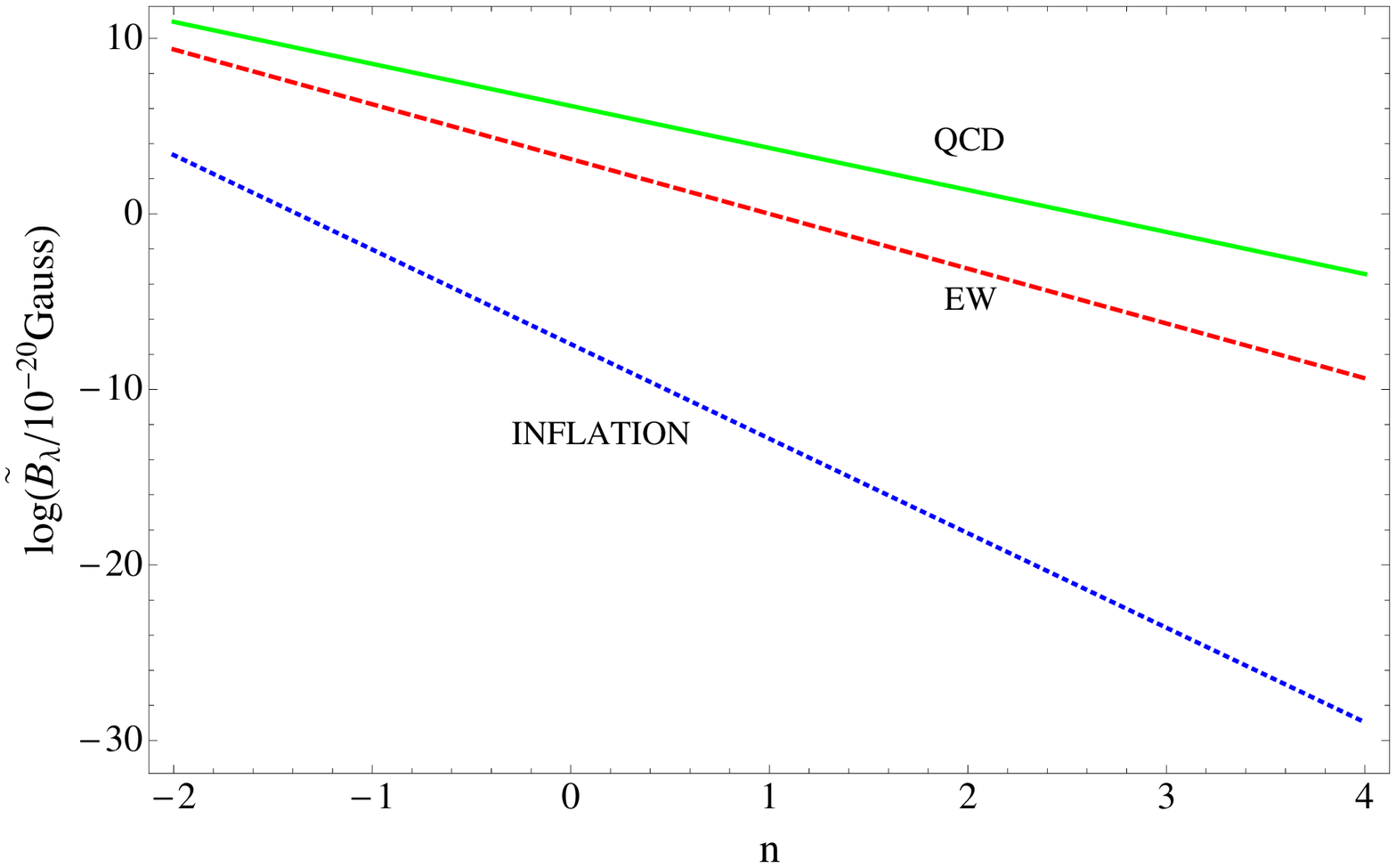}\\
\caption{Upper bounds on the comoving amplitude of a primordial magnetic field from GW production as a function of $n$, for $\ti\la=0.1$ Mpc, for 
a field generated at inflation (blue, dotted), and at a phase transition (EW: red, dashed, QCD: green, solid). In these last two cases, the generation is causal, consequently only $n\geq 2$ is allowed.}
\label{f:limits}       
}  

More stringent bounds on the amplitude of a causally produced magnetic field can be obtained by imposing simply that the energy density of the magnetic field cannot overcome 10\% of the total energy density in radiation at generation time: $$\ti \Om_B^*\leq 0.1\,\Om_{\rm rad}\ .$$ Comparing this last inequality with Eq.~(\ref{OmBEWlim}), one sees that in the latter the factor multiplying $\Om_{\rm rad}$ is 5 orders of magnitude larger than 0.1 (with the usual values for the parameters). This shows that, in the causal case, the conversion of magnetic energy density into GW energy density, although quite efficient, is not at all complete. For example, using $\ti \Om_B^*\leq 0.1\,\Om_{\rm rad}$, the bound on an helical magnetic field becomes $\ti B_{0.1{\rm Mpc}}\lesssim 5\times 10^{-26}$ Gauss for the EW phase transition. In the inflationary case, on the other hand, the bounds are not modified, since the conversion into GW is much more efficient (\cf~the pre-factor in Eq.~(\ref{OmBINFlim})). 

However, accounting for GW production seems to us more model independent. Once GWs are generated they do not interact with the cosmic fluid and simply redshift with the evolution of the universe. We are therefore sure that any GW energy density sourced before nucleosynthesis is still present at that time and must respect the nucleosynthesis bound. On the other hand, magnetic energy density can be dissipated or converted into other forms of energy during the evolution of the universe. We could therefore invoke the extreme scenario in which a magnetic field is formed in the very early universe with amplitude higher than the presumed radiation energy density at that time. The magnetic energy can subsequently be transformed into other forms of energy in such a way that it satisfies the nucleosynthesis bound at nucleosynthesis. The main motivation to consider GW production is to obtain a bound which is safe from these exotic, but in principle possible, scenarios.

\section{Conclusions}\label{s:con}
In this paper we have derived new upper bounds on the amplitude of primordial magnetic fields. 
We have considered helical magnetic field power spectra, which evolve via inverse 
cascade transferring power from small to large scales. 

For the non-helical case, upper bounds on the magnetic field amplitude on the cosmologically relevant scale $\ti\la \simeq 0.1$ Mpc have been derived in previous analyses \cite{Caprini:2001nb,Caprini:2006jb}. These bounds apply to magnetic fields generated before nucleosynthesis, for which the correlation scale at the moment of generation is $\ti L_*\ll \ti \la$. If the magnetic field spectrum is blue $n> -3/2$, the peak of the energy density per logarithmic scale sits at $\ti L_*$, then the amplitude at the scale $\ti \la$ is very constrained: $$\ti B_\la \simeq \ti B_{L_*}(\ti L_*/\ti\la)^{(n+3)/2} \ll \ti B_{L_*}$$.

On the other hand, if an inverse cascade is active, the power at $\ti L_*$ is moved to the 
larger correlation scale $\ti L(\tau)$, following the evolution law $\ti L(\tau)=\ti L_*\tau^{2/3}$. At the end of the inverse cascade process, we have seen that the magnetic field on the scale $\ti\la$ is finally 
\be
\ti B_\la \simeq \ti B_{L_*}(\ti L_*/\ti\la)^{(n+3)/2}\tau_{\rm fin}^{(n+2)/3}
\ee 
(naively one might expect a scaling like $\tau_{\rm fin}^{(n+3)/3}$, but some of the initial amplitude is lost during the inverse cascade process, 
so that $\ti \Om_B (t)\ti L(t) =$ constant. This reduces the growth of $\ti B_\la$ by a 
factor $\tau_{\rm fin}^{1/3}$). 

The strong limits from magnetic fields which obey a direct cascade are 
therefore mitigated in the helical case by a factor 
\bea
\tau_{\rm fin}^{(n+2)/3} &=& \lp[ \frac{2v_L}{\ep} \,\frac{T_*}{T_{\rm fin}} \lp( \frac{g_*}{g_{\rm fin}} \rp)^{1/6} \rp]^{(n+2)/3} \vspace{6pt}\\
 &\simeq& \left\{ \begin{array}{ll} \nn
 	5.3 \times 10^7 & \mbox{for the EW transition, } 
  		\\ &   n=2~,~   T_*=100{\rm GeV}\\
	2\times 10^4 & \mbox{for the QCD transition, } \\ &  n=2~,~ 
   		T_*=100{\rm MeV}\\
	9.3 & \mbox{for inflation, }  \\ &  n=-1.8~,~ T_*= 10^{14} {\rm GeV}\\
	4.8 \times 10^9 & \mbox{for inflation, } \\ &  n= 0 ~,~ 
		T_*= 10^{14} {\rm GeV}\,.
\end{array} \right.
\eea
In this paper we have only considered spectral indexes $n>-2$, since Ref.~\cite{Campanelli:2007tc} does not analyse smaller spectral indexes, and numerical simulations have always chosen either $n=2$ \cite{Christensson:2000sp}, or $n=0$ \cite{Banerjee:2004df}. Consequently, we do not know whether for a red spectrum with $n\le -2$ the inverse cascade is still active, or whether the limits are those of the direct cascade. For spectral indexes close to $ n=-2$ the above mitigating factor is small, even if the magnetic field is generated in the very early universe. On the other hand, for blue magnetic fields generated at the QCD phase transition, the inverse cascade is not very 
efficient, since turbulence anyway stops after $e^+e^-$ annihilation: therefore   
$\tau_{\rm fin}$ is not very large.

In Table \ref{tab} we summarise our results. We give the upper bounds on the magnetic field amplitude obtained both accounting for GW production and imposing that the magnetic energy does not overcome 10\% of the radiation energy, for helical and non-helical magnetic fields smoothed on scales of 0.1 Mpc and 1 kpc. 
The upper bounds for a non-helical magnetic field are in agreement with those given in \cite{Caprini:2001nb,Caprini:2006jb}.

We have found that only red magnetic field spectra from inflation or helical fields from the
QCD phase transition can have the amplitude of
$B_{0.1{\rm Mpc}} \gsim 10^{-22}$ Gauss which is necessary for amplification by a dynamo mechanism up to the observed $\mu$Gauss 
field~\cite{BS}. Especially, the well motivated 
helical fields from the EW phase transition are still too constrained, even after the inverse cascade. 

This leads us to the conclusion that the observed magnetic fields in 
galaxies and clusters have either not been seeded by primordial fields, or these
primordial fields have been produced during inflation and have a red 
spectrum $n\lsim -1.8$, or they have been produced during the QCD phase transition.
In this latter case it is crucial that these QCD fields be helical because the
boost by the factor of about 20000 is absolutely needed, while for red ($-2<n<-3/2$)
inflationary fields the inverse cascade is not relevant.

To evade this conclusion one can argue that magnetic fields coherent on a smaller scale, of about 1 kpc, and with the required amplitude of $10^{-22}$ Gauss are sufficient to give rise to the fields observed today in galaxies and clusters \cite{Banerjee:2004df}. If this is so, then the bounds derived here are relaxed in such a way that also helical fields from the EW phase transition can have a sufficiently high amplitude (however, for non-helical fields this is still not enough -- see Table \ref{tab}).

\begin{table}
\begin{center}
\begin{tabular}{|l|l|l|l|l|l}
\hline
\multicolumn{5}{|c|}{  }\\
\multicolumn{5}{|c|}{GW limits}\\ [6pt]
\hline\hline
& \multicolumn{2}{|c|}{ } & \multicolumn{2}{|c|}{  } \\
  & \multicolumn{2}{|c|}{helical} & \multicolumn{2}{|c|}{non-helical} \\ [6pt]
  \hline
  & $\la=0.1$Mpc & $\la=1$kpc & $\la=0.1$Mpc & $\la=1$kpc \\ [6pt]
  \hline
  EW & $8\times 10^{-24}$G &  $\bf 8\times 10^{-19}$ {\bf G} & 
$10^{-31}$G & $ 10^{-26}$G \\ [2pt]
  QCD & $\bf 2\times 10^{-19}$ {\bf G} &  $\bf 2\times 10^{-14}$ {\bf 
G} & $10^{-23}$G & $\bf  10^{-18}$ {\bf G} \\ [2pt]
   Infl. $n=-1.8$ & {\bf $\bf 2\times 10^{-18}$G} &  $\bf 3\times 
10^{-17}$ {\bf G} & $\bf 2\times10^{-19}$
{\bf G} & $\bf 3\times 10^{-18}$ {\bf G} \\ [2pt]
    Infl. $n=0$ & $4\times 10^{-28}$G &  $4\times 10^{-25}$G & 
$8\times10^{-38}$G & $8\times 10^{-35}$G \\ [6pt]
   \hline\hline
   \multicolumn{5}{|c|}{  }\\
\multicolumn{5}{|c|}{limits from $\Om_B^* < 0.1$}\\ [6pt]
\hline\hline
EW & $5\times 10^{-26}$G &  $\bf 5\times 10^{-21}${\bf G} & 
$6\times10^{-34}$G & $6\times 10^{-29}$G \\ [2pt]
  QCD & $\bf 10^{-21}${\bf G} &  $\bf 10^{-16}${\bf G} & 
$6\times10^{-26}$G & $\bf 6\times 10^{-21}${\bf G} \\ [2pt]
   Infl. $n=-1.8$ & $\bf 2\times 10^{-18}${\bf G} &  $\bf 3\times 
10^{-17}${\bf G} & $\bf 2\times10^{-19}${\bf G} & $\bf 3\times 
10^{-18}${\bf G} \\ [2pt]
    Infl. $n=0$ & $4\times 10^{-28}$G &  $4\times 10^{-25}$G & 
$8\times10^{-38}$G & $8\times 10^{-35}$G \\ [2pt]
   \hline
\end{tabular}
\end{center}
\caption{\label{tab} This table summarises the upper bounds for the magnetic field amplitude 
averaged over the scales $\la=0.1$ Mpc
and $\la=1$ kpc, for the different generation epochs discussed in the paper. Here we present the limits for maximally
helical as well as non-helical fields. In the four top rows we give 
the limits from   the production
of GWs while in the lower part of the table we present 
the limits coming from
the requirement that the magnetic field contribution be always 
subdominant: more precisely we require $\Om_B^* < 0.1$. The values which may be sufficient for dynamo 
amplification are given in boldface. }
\end{table}

\vspace*{0.8cm}
\noindent
{\large{\bf Ackowledgements}}\\
We thank K. Jedamzik, T. Kahniashvili, G. Servant, K. Subramanian and T. Vachaspati for useful discussions. EF thanks CEA-Saclay for hospitality. This work is supported by the Swiss National Science Foundation. 

\vspace*{0.8cm}
\noindent
{\large{\bf Appendix}}
\appendix

\section{The end of turbulence and the dissipation scale}
\subsection{The end of turbulence \label{A:Tfin}}

The turbulent phase ends when the Reynolds number on the  scale of energy 
injection becomes of order unity~\cite{LL}
\bea
\Re(L(T_{\rm fin}),T_{\rm fin})=\left.\frac{v_L \,L}{\nu}
  \right|_{T_{\rm fin}}\simeq 1 \, .
\label{ReL=1}
\eea
Here $L$ denotes the {\em physical} energy injection scale, $\nu$ is the 
kinetic viscosity, and $v_L$ is the eddy velocity on the scale $L$.  We assume
that in the MHD cascade kinetic and magnetic energy have the same  time 
evolution (equipartition). Substituting 
$L(t)= \ti L(t)a(t)=\ti L(t)\,(T_0/T(t))\, (g_0/g_{\rm eff}(T))^{1/3}$, 
we find ($g_{\rm eff}(T_*)\equiv g_*$),  
\be
\Re(L(T),T)=R_*\frac{T_*}{T}\lp(\frac{g_*}{g_{\rm eff}(T)}\rp)^{1/3} \frac{\nu_*}{\nu(T)}\,\tau^{\al} \, , \qquad R_* \equiv 
  \Re(L_*,T_*)   ~,
\label{ReL}
\ee
where the power $\al$ represents the evolution of the product $\ti L(t)v_L(t)$ 
and we use $v_L(t)\propto \sqrt{\ti\rho_B}$. With
Eqs.~(\ref{e:EBsd}, \ref{e:Lsd}) and (\ref{Einverse}, \ref{Linverse}) we obtain
\be \label{e:alpha}
\al= \left\{ \begin{array}{ll}
 -(n+1)/(n+5) & \mbox{for non-helical fields (normal cascade),} \\
   1/3  &\mbox{for helical fields (inverse cascade).} 
\end{array} \right.
\ee

The kinetic viscosity is approximately given by the mean free path of the
 particle with the weakest interaction~\cite{Wei}, 
 $\nu\simeq \ell_{\rm mfp}/5$. 
During the early radiation dominated phase, neutrinos determine the viscosity 
until they decouple at $T\simeq 1$ MeV, when photon viscosity sets in. For 
$T>1\,$MeV the mean free path of 
the neutrinos is given by
\be\label{e:ellnu}
\ell_{\rm mfp}^{(\nu)}\simeq \frac{1}{3\,G_F^2 \,T^5}\,, \qquad 
                     1\,\mbox{MeV} < T < 100\,\mbox{GeV}~,
\ee
where $G_F=(293{\rm GeV})^{-2}$ is the Fermi coupling constant. Below $1$ MeV 
we have to consider
the photon mean free path which can be approximated by Thomson scattering
\be \label{e:thom}
\ell_{\rm mfp}^{(\gamma)}\simeq \frac{1}{\sigma_T\,n_e}\simeq 
\frac{3\, m_e^2\,m_p}{8\pi e^4\,\Omega_b\,\rho_c}\left(\frac{T_0}{T}\right)^3\,,
    \qquad   0.3\mbox{ eV} < T < 1\mbox{ MeV} ~,
\ee
where $\sigma_T$ is the Thomson cross section, $n_e$ is the electron density, $\Om_b$ is the baryon density parameter  
and we neglect the short period of time during which electrons are 
still relativistic, after neutrino decoupling but before electron positron 
annihilation. The lower
limit in (\ref{e:thom}) comes from recombination, when the electron density 
drops sharply and photons decouple.
The situation also changes at very high temperature, when the EW 
symmetry is restored, $T\gsim 100$GeV. Then the coupling constant is nearly 
independent of temperature, and the relativistic mean free path is of the 
order of the inverse temperature \cite{arnold}
\be\label{e:ellhi}
\ell_{\rm mfp}\simeq \frac{22}{T}\,, \qquad T> 100\,\mbox{GeV} ~ .
\ee 

We first calculate the Reynolds number at the time of generation and at the 
initial correlation scale of the magnetic field, in order to confirm that a 
turbulent MHD phase is indeed present. With (\ref{ReL}) we obtain
\be
R_* =v_L\,\epsilon \,\frac{t_\in}{\nu_*}\,\frac{T_0}{T_*} \lp( \frac{g_0}{g_*} \rp)^{1/3} \,,
\ee
where we have identified the temperatures $T_\in \simeq T_*$. 
Furthermore, we use $\nu = \ell_{\rm mfp} /5$ and 
\bea
t_\in &=& \frac{1}{a_\in H_\in} \simeq 0.5\,g_*^{-1/6}
  \frac{m_{\rm Pl}}{T_*T_0}\,, \nonumber \\
R_* &\simeq&  \frac{3 \, v_L\,\epsilon}{\sqrt{g_*}} 
 \, \frac{m_{\rm Pl}}{T_*^2\ell_{\rm mfp}(t_*)} \,.
 \label{e:Regen}  
\eea
We consider the two situations, $1\,{\rm MeV} \le T_*\le 100\,{\rm GeV}$ and 
$T_*>100$ GeV, where the expressions (\ref{e:ellnu}) and  (\ref{e:ellhi}) 
for the mean free path give
\bea
R_*  \simeq& \frac{9 v_L\,\epsilon}{\sqrt{g_*}} 
m_{\rm Pl}T_*^3G_F^2\,, & 1\,{\rm MeV} \le T_*\le 100\,{\rm GeV} \\
R_* \simeq&  \frac{3}{22} \frac{v_L\,\epsilon}{\sqrt{g_*}} 
 \, \frac{m_{\rm Pl}}{T_*}\,, & T_*>100 \mbox{ GeV} \,.
\eea

Let us start by considering 
the generation of turbulence during the EW phase transition at 
100 GeV. Setting $T_*=100\,$ GeV, $\ep=0.01$ and $v_L^2=0.2$, with 
$g_* \simeq 100$ we obtain $R_* \simeq 10^{13}$. The 
corresponding parameters for the QCD transition at $T_*=100$ MeV and $g_* \simeq 10$ yield
 $R_* \simeq 10^{4}$. Both Reynolds numbers are much larger than 
one so that we can be certain that a first order phase transition will 
induce turbulence. To determine when turbulence terminates we use 
Eqs.~(\ref{ReL}) and (\ref{e:ellnu}), setting
\be
1 = \Re(L(T_{\rm fin}),T_{\rm fin}) = R_{*}\tau^\al \lp(\frac{g_*}{g_{\rm eff}(T_{\rm fin})}\rp)^{1/3}   \lp(\frac{T_{\rm fin}}{T_*}\rp)^4 \simeq 
R_*\left(\frac{2 v_L}{\ep}\right)^\al
    \lp(\frac{T_{\rm fin}}{T_*}\rp)^{4-\al} \,.
\ee
For the last equal sign we have used
$t_L^* =\ti L_*/(2v_L)= \ep t_\in / (2v_L)$,  
$ \tau \simeq t/t_L^*$ and we have approximated $t_{\rm fin}/t_{\rm in}\simeq 
T_*/T_{\rm fin}$. This corresponds to neglecting changes in the number of 
effective relativistic degrees of freedom. 

For direct cascade with $n=2$, hence $\al=-3/7$, we obtain 
$T_{\rm fin}\simeq 200$ MeV for the EW phase transition and 
$T_{\rm fin}\simeq 20$ MeV for the QCD phase transition.

In the helical case with inverse cascade $\al=1/3$, turbulence is maintained 
longer and we find
$T_{\rm fin}\simeq 21$ MeV for the EW phase transition and 
$T_{\rm fin}\simeq 5$ MeV for the QCD phase transition.

For generation of magnetic fields and turbulence at the end of inflation $T_*=10^{14}$ GeV, no causality restriction holds and we choose 
$\epsilon=v_L = 1$. As long as $\nu(T)\propto \ell_{\rm mfp}\simeq 22/T$, 
the Reynolds number at the correlation length $L(T)$ evolves like 
$(t/t_{\rm in})^\al\simeq (T_*/T)^\al$ according to 
Eq.~(\ref{ReL}) (where we neglect changes in the effective number of relativistic degrees of freedom). 
At $T_*\simeq 10^{14}$ GeV, with $g_*\simeq 200$
we have $R_* \simeq 0.01\,m_{\rm Pl}/T_* \simeq 10^3$. As time evolves, 
the Reynolds number  at $L(T)$ decays only in the non-helical case 
if $n>-1$, so that $\al=-(n+1)/(n+5)<0$. In the helical case and for $n<-1$ 
the Reynolds number $\Re(L(T),T)$ grows as the temperature drops. Once 
$T=T_{ew}=100$ GeV is reached, the viscosity  $\nu$ starts decaying rapidly, like $T^{-5}$, and 
the Reynolds number then decreases. 

We consider three cases\\
 {\bf i) Direct cascade with} $n=0$, $\al=-1/5$
$$\Re(L(T_{ew}),T_{ew})=R_*(T_*/T_{ew})^{\al} \simeq 6\,,$$
so $T_{\rm fin} \simeq 100$ GeV.\\
{\bf ii) Direct cascade with} $n=-3/2$, $\al=1/7$  \\
In this case 
$$\Re(L(T_{ew}),T_{ew}) = R_*(T_*/T_{ew})^{\al} \simeq 9\times 10^4\,, $$
 and
$$ 
T_{\rm fin} = T_{ew}\Re(L(T_{ew}),T_{ew})^{\frac{-1}{4-\al}} \simeq 5\,{\rm GeV}~ .
$$
{\bf iii) Inverse cascade}, $\al=1/3$   \\
In this case 
$$\Re(L(T_{ew}),T_{ew})=R_*(T_*/T_{ew})^{\al} \simeq 2 \times10^7\,, $$
 and
$$ 
T_{\rm fin} = T_{ew}\Re(L(T_{ew}),T_{ew})^{\frac{-1}{4-\al}} \simeq 1\,{\rm GeV}~ .
$$

We draw the important conclusion that in all cases the MHD 
turbulent phase always lasts for many Hubble times before the total kinetic 
energy is dissipated \cite{CDS2}. 

\subsection{The dissipation scale\label{A:ldiss}}
In the previous subsection we have considered the energy injection scale $\ti L(T)$ and determined  
first that turbulence is present on this scale, and second when turbulence ends ({\it i.e.} when the entire Kolmogorov range is dissipated). 
Now we want to know, for a given fixed time 
$t$ (or temperature $T$), what is the scale below which kinetic energy is 
dissipated. This defines the comoving dissipation scale $\ti l_{\rm diss}(T)$.
The function $\ti l_{\rm diss}(T)$ can 
be found considering that, on scales smaller than this scale,  
viscosity dominates, therefore there is no turbulence. 
Thus, $l_{\rm diss}$ corresponds to the physical scale at which the Reynolds number is equal to 1,
\be\label{e:defdis}
  \Re(l_{\rm diss},T)=\frac{v_l\, l_{\rm diss}}{\nu} \sim 1\,.
\ee
Here $\nu$ is the kinetic viscosity as in the previous section,
and $v_l$ is the eddy velocity at the dissipation scale.
We determine $\ti l_{\rm diss}(T)$ only for the helical case which is our 
main interest in this paper (see \cite{CDS2} for the non-helical one).  In the inertial range the turbulent eddy
velocity obeys a Kolmogorov spectrum so that~\cite{Caprini:2006jb} 
\be
v_l = v_L\lp(\frac{\ti  l_{\rm diss}}{\ti L(T)} \rp)^{1/3}\, .
\ee
Now we use $\Re(L(T),T) =v_L(T) L(T)/\nu(T)$ so that, from Eq.~(\ref{ReL}) 
neglecting changes in the number of relativistic degrees of freedom, we find 
\be\label{Relt} 
\Re(l_{\rm diss},T)= \Re(L(T),T)\lp(\frac{\ti  l_{\rm diss}}{\ti L(T)}\rp)^{4/3}
  \simeq R_*\tau^\al\frac{T_*}{T} \frac{\nu_*}{\nu} \lp(\frac{\ti l_{\rm diss}}{\ti 
   L(t)}\rp)^{4/3} \,. 
\ee

Hence $\Re(l_{\rm diss},T)=1$ yields
\be	\label{ldiss}
	  \ti l_{\rm diss}(T)\simeq \ti l_{\rm diss}^* \,\tau^{5/12}  \lp(\frac{T}{T_*} \frac{\nu(T)}{\nu_*} \rp)^{3/4}  ~,
\ee
where we define $\ti l_{\rm diss}^* \equiv \ti L_* / R_*^{3/4}$.
For the last equal sign we have used the behavior of the correlations
scale with $\tau = (t-t_\in)/t_L^*$ as $\tau^{2/3}$, (\ref{Linverse}), and $\al=1/3$ for the helical case.
 
The evolution of both the correlation length $\ti L(T)$ and the 
dissipation scale $\ti l_{\rm diss}(T)$ are compared in 
Fig.~\ref{fig:ldiss} for the EW phase transition. Turbulence stops 
roughly when the two curves cross. We call this time $t_{\rm fin}^{(1)}$.

Finally, we have to take into account that for a given comoving scale,
 $\ti l=2\pi/k$ the Reynolds number can become unity long before
the end of turbulence.
The time at which turbulence on the scale $\ti l$ is 
dissipated is denoted $t_{\rm fin}^{(2)}(k)$ and it is defined by 
$$\ti l=2\pi/k = \ti l_{\rm diss}(t_{\rm fin}^{(2)}(k) ) \,.$$

Let us first consider $T_*\le T_{ew}=100$ GeV, so that for all times of interest
the kinetic viscosity behaves as $\nu \propto T^{-5}$. This leads to
\be\label{e:tk2EW}
   k=\frac{2\pi}{\ti l_{\rm diss}(t_{\rm fin}^{(2)}(k))} \quad 
\Rightarrow \quad \frac{t_{\rm fin}^{(2)}(k)}{t_*} \simeq  
\lp[  \frac{1}{K_*}\frac{\ti L_*}{\ti l_{\rm diss}^*}
   		\lp( \frac{\epsilon}{2v_L}\rp)^{5/12} \rp]^{12/41}  ~.
\ee
Here again, we neglect a possible difference in $g_{\rm eff}$ between 
$t_*$ and $t_{\rm fin}^{(2)}$ and we set 
$\tau_{\rm fin}^{(2)} \simeq t_{\rm fin}^{(2)}2v_L/(\ep t_\in)$.
We also use $R_* =(\ti L_*/\ti l^*_{\rm diss})^{4/3}$.

The situation is somewhat more complicated for generation temperatures $T_*>T_{ew}$: until
$T_{ew}$ the kinematic viscosity decays roughly like $\nu\propto 1/T$. We therefore 
have to distinguish between scales which are damped at temperatures above 
$T_{ew}$ and those which are damped below. Since we are in this situation only 
for the inflationary case, we set $\ep=v_L= 1$ for this case.
We then obtain
\bea \label{e:tk2inf}
t_{\rm fin}^{(2)}(k) \simeq \left\{ \begin{array}{ll}
 	\frac{t_*}{2} \left(K_*^{-1}\frac{\ti L_*}{\ti l_{\rm diss}^*}\right)^{12/5} &
	 	\mbox{ for } K_* \mbox{ such that } t_{\rm fin}^{(2)}(k) <t_{ew}\, , \quad
       	K_* > K_{ew}^*~,  \vspace{8pt}\\
  	t_{ew}\left(\frac{K_{ew}^*}{K_*}\right)^{12/41} 2^{-5/41} & \mbox{ for } K_*<K_{ew}^* \,.
\end{array} \right.
\eea
Here $K_{ew}^*$ is the value of $K_*$ for which turbulence terminates at
$T_{ew}$,
\bea
\label{e:Kew*}
K^*_{ew} \simeq \left(\frac{T_{ew}}{2T_*}\right)^{5/12}\frac{\ti L_*}{\ti l_{\rm diss}^*}
    	\,.
\eea

When applying these formulas for the calculation of GWs, 
we must choose the true final time given by
\be
   t_{\rm fin} = {\min} \lp[t_{\rm fin}^{(1)}, t_{\rm fin}^{(2)}(k) \rp]  ~.
\ee
For turbulence from the EW phase transition and from inflation this function is plotted in Fig~\ref{fig:tfin} .
At $t_{\rm fin}^{(1)}$ turbulence is dissipated on all scales 
$\ti l \leq\ti L(t_{\rm fin}^{(1)})$ which is the scale of the largest eddies:   
therefore, the entire Kolmogorov range is dissipated.

\section{The equal time correlator and other integrals}

\subsection{ $\ti\Pi_B(k,t)$ \label{A:Pi}}
According to Eq.~(\ref{e:Pitt}) we can write the equal time correlator
as the following integral
\bea
  \ti\Pi_B(k,t)&=& \frac{\NN_2}{2\pi} \ti L^3(t) 
    \ti\rho_B^2(t)\lp[ I_1(k,t)+ 4 I_2(k,t) \rp] = 
       \frac{\NN_2}{2\pi} \ti L^3(t) \ti\rho_B^2(t)  I(k,t)  ~,
  \\
  I_1(k,t)&=&  \int_0^{\ti L(t)/\ti l_{\rm diss}(t)} dQ 
 	\frac{ Q^{n+2}}{(1+Q^2)^{(2n+7)/4}} \times
  	\nn \\ &&
 \int_{{\rm max}\lp(-1; \frac{K}{2Q} +\frac{Q}{2K}- 
       \frac{\ti L^2}{2KQ \ti l_{\rm diss}^2} \rp)}^1 
  d\gamma (1+\gamma^2) \frac{  x^{n-2}}{(1+x^2)^{(2n+7)/4}}
         \times\nn \\ &&
	[2K^2+(1+\gamma^2)Q^2-4\gamma KQ]  ~,
  \\
  I_2(k,t)&=& \int_{\ti L(t)/(2\pi t)}^{\ti L(t)/\ti l_{\rm diss}(t)} dQ 
 	\frac{  Q^{n+2}}{(1+Q^2)^{(2n+7)/4}} \times
  	\nn \\ &&
  \int_{{\rm max}\lp(-1; \frac{K}{2Q} +\frac{Q}{2K}- 
   \frac{\ti L^2}{2KQ \ti l_{\rm diss}^2} \rp)}
  ^{{\rm min}\lp(1; \frac{K}{2Q} +\frac{Q}{2K}- \frac{\ti L^2}{2KQ (2\pi t)^2} \rp)} 
  	d\gamma ~ \frac{  x^{n-1}}{(1+x^2)^{(2n+7)/4}}
         \gamma (K-Q\gamma)    ~,
\eea
where we have used $\alpha = (k-\gamma q)/\sqrt{k^2+q^2-2\gamma kq}$ and we 
have set\\ $x \equiv 
\sqrt{K^2+Q^2-2\gamma KQ}$ and  $Q(t) \equiv q \ti L(t)/(2\pi)$. We have also 
introduced the constant $\NN_2$ defined by
\[
 \NN_2 \equiv  \lp[\frac{\Gamma\lp( \frac{2n+7}{4} \rp)}
{\Gamma\lp( \frac{1}{4} \rp)\Gamma\lp( \frac{n+3}{2} \rp)} \rp]^2 =\left\{
\begin{array}{ll}
 0.11 & \mbox{ for } n=2~, \\
 0.08 & \mbox{ for } n=0 ~,\\
 0.05 & \mbox{ for } n=-3/2 ~,\\
 0.04 & \mbox{ for } n=-1.8 ~. \end{array} \right.
 \]

We have performed the double integrals $I_j(k,t)$ numerically 
for different values of the spectral index $n>-2$. 
In the numerical integration we neglect the time dependent cutoff in the 
above integrals and consider only the time dependence given by
$K(t)=K_* \tau^\beta$, with $K_*\equiv k \ti L_* /(2\pi)$, 
$\tau \equiv [(t-t_{\rm in})/t_L^*]$ and $\beta=2/(n+5)$ for the selective decay 
regime while $\beta=2/3$ in the inverse cascade phase. In general we find that 
the antisymmetric contribution $I_2$ is negative (as it should be~\cite{Caprini:2003vc}), and 
negligible with respect to the symmetric contribution, namely $I_1\ll |I_2|$.

\FIGURE[ht]{ 
\epsfig{width=10cm, file=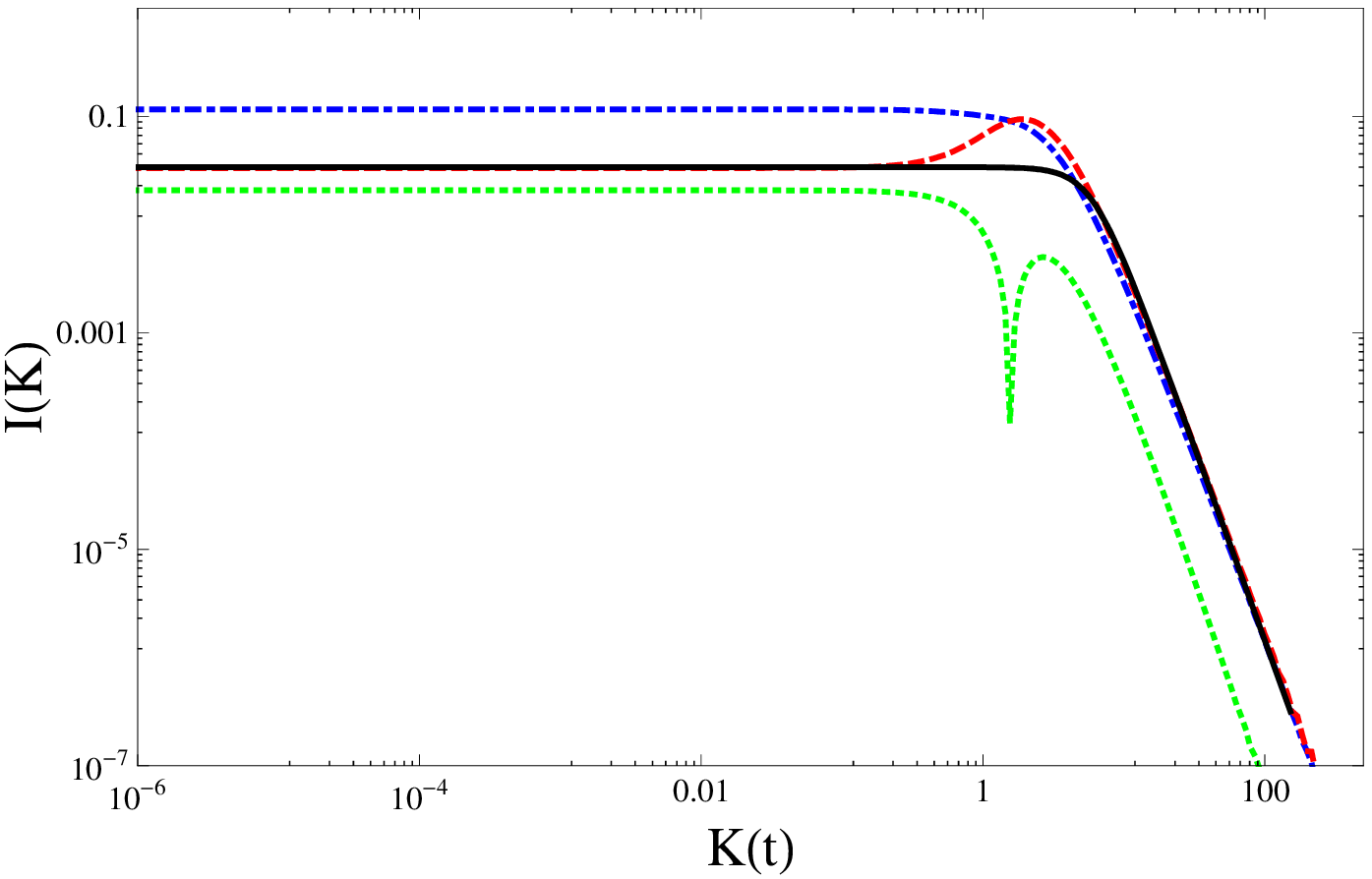}\\
\caption{Symmetric part $I_1(K)$ (blue, dash-dotted line) and absolute value of
the antisymmetric part $|I_2(K)|$ (green, dotted line) of the anisotropic 
stresses for a blue magnetic field with $n=2$. The sum of the two parts 
$I(K)=I_1+4I_2$ (red, dashed line)  and the fit (black, solid line) are also 
shown. }
\label{fig:piBlueK}       
}  
\FIGURE[ht]{ 
\epsfig{width=10cm, file=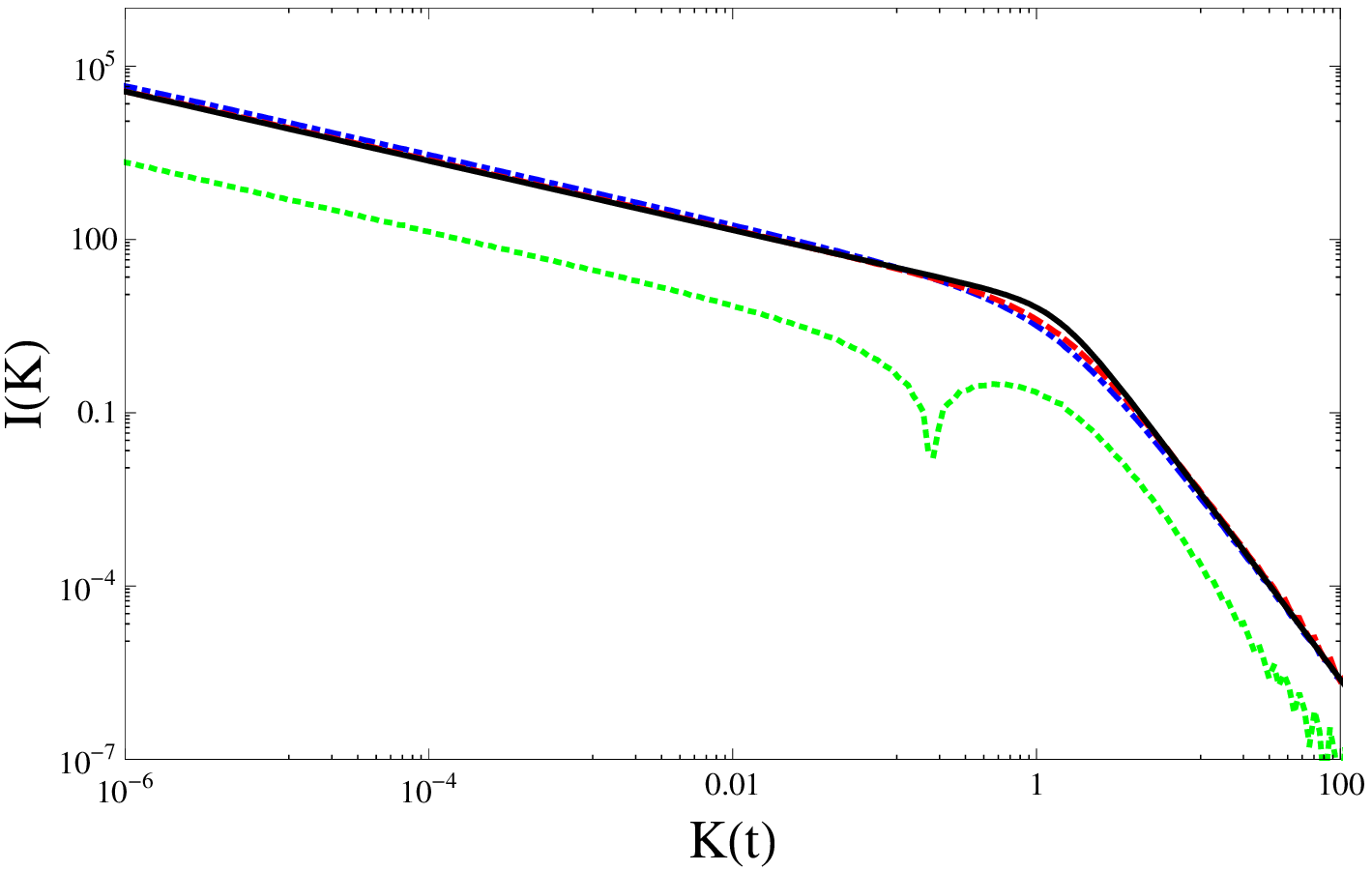}\\
\caption{Symmetric part $I_1(K)$ (blue, dash-dotted line) 
and absolute value of the antisymmetric part $|I_2(K)|$ (green, dotted line) 
of the anisotropic stresses considering now a red magnetic field with 
$n=-1.8$. We plot also the sum of the two parts $I(K)$ (red, dashed line)  and 
the fit (black, solid line). }
\label{fig:piRedK}       
}  

  First we consider a blue magnetic field spectrum which 
  is characterized by $n>-3/2$. The result of the 
  numerical integration for $n=2$ , which corresponds to a causal 
magnetic field, is shown in Fig.~\ref{fig:piBlueK}. 
The integral $I(K)$ can be approximated by 
  the following analytical expression,
  \be
    \label{PiBlueK}
    I(K)\simeq \frac{0.034}{1+\lp(\frac{K}{12}\rp)^4+\lp(\frac{K}{6}\rp)^{7/2}  }  ~.
  \ee
  With this approximation the equal time correlator $\ti\Pi_B(K,t)$ can be written as
  \be
  	\label{A:pi4K}
    \ti\Pi_B(K,t)\simeq \frac{\NN_2}{2\pi} \ti L^3(t) \ti\rho_B^2(t)  
    \frac{0.034}{1+\lp(\frac{K}{12}\rp)^4+\lp(\frac{K}{6}\rp)^{7/2}  }  ~.
  \ee

  For a red magnetic field spectrum with $n<-3/2$ we find a somewhat 
different power spectrum which is shown in Fig.~\ref{fig:piRedK} for the case 
$n=-1.8$. The integral can be approximated by
\be
    I(K)\propto  \frac{K^A}{1+(K/C)^B  }   ~,
  \ee
  where the constants $A$ and $B$  are given by
 $A=2n+3<0$ and $B= A+7/2 = 2n+13/2$ and $C$ varies with $n$ and has been chosen to be equal to $1.4$ in order to fit the numerical result. 
 We recover the
asymptotic behavior for the anisotropic stress of a red magnetic field 
found in~\cite{Caprini:2001nb},~\cite{Caprini:2006jb}:
  \bean
  && 	I(K)\simeq K^{2n+3} \qquad {\rm for } \quad K \ll 1  ~,
  \\
  && 	I(K)\simeq K^{-7/2}   \qquad {\rm for } \quad K \gg 1  ~.
  \eean 
This explains also the higher amplitude of $I(K)$ in the red case compared to
the one obtained for a blue magnetic field: we have an infrared 
divergence for small values of $Q$. \\
Finally for $n=-1.8$ we find
  \be
  	\label{A:pi0.2K}
  \ti\Pi_B(K,t)\simeq \frac{\NN_2}{2\pi} \ti L^3(t) \ti\rho_B^2(t) 
\frac{(K/40)^{-3/5}}{1+(K/1.4)^{29/10}  }   ~. 
  \ee

\subsection{GW integrals \label{A:gwint}}
To calculate the integrals~(\ref{drhoGW})
we use Eqs.~(\ref{pi4K}),~(\ref{piRed}) for the time evolution of the magnetic 
anisotropic stress and we use the integration variable $\tau$
	\[
	      \tau=\frac{t'-t_{\rm in}}{t_L^*}  ~.
	\]
We can then write the GW energy spectrum as
\bea
\frac{ d\rho_{\rm GW}(k,t)}{d\log k} &\simeq& 
  \frac{2 G }{\pi a^4(t) } \frac{ \NN_2}{2\pi}
	\frac{(\ti\rho_B^*)^2 }{H_0^2  \Omega_{\rm rad}}
	 (2\pi)^3 K_*^3  \lp\{  \lp[I_{c,\,I}(k)+I_{c,\,II}(k)\rp]^2  \rp.
	\nn \\ &&  \lp.
	+ \lp[I_{s,\,I}(k)+I_{s,\,II}(k)\rp]^2  \rp\}  ~, 
\eea	
which yields the following expression for the present density parameter of 
GWs
\bea
  \frac{d\Omega_{\rm GW}(k,t_0)}{d\log k} &\simeq& 3\,
 \NN_2 \frac{(\ti\Omega_B^*)^2}{\Omega_{\rm rad}} \II_{\rm GW}(K_*)  ~,
	\nn  \\
	\II_{\rm GW}(K_*) &\equiv& K_*^3 
			\lp\{  \lp[I_{c,\,I}(K_*)+I_{c,\,II}(K_*)\rp]^2  
		+ \lp[I_{s,\,I}(K_*)+I_{s,\,II}(K_*)\rp]^2  \rp\}  ~. \label{app:rhoGW}
\eea	
The four integrals  above distinguish the two different phases of the 
inverse cascade, namely the first one where the magnetic energy density is growing linearly up to its maximum value $\ti\rho_B^*$ ($t_\in \leq t <t_*$) and the second one where $\ti\rho_B(t)$ decays as $\tau^{-2/3}$ ($t \geq t_*$).

For the EW phase transition, $n=2$ we have for  $t_{\rm in}  \leq t < t_*$
\bea
   I_{c,\,I}(K_*) &=&     \int_0^1 
 	 d\tau \frac{\sqrt{0.034}~\tau^2}{\lp(\tau+\frac{2v_L}{\epsilon}\rp) 
	 \sqrt{1+\lp(\frac{K_*\tau^{2/3}}{12}\rp)^4+\lp(\frac{K_*\tau^{2/3}}{6}\rp)^{7/2}}}  	
	 \times \nn \\ && 
 \cos\lp[2\pi  K_*\lp(\frac{\tau}{2v_L} +\frac{1}{\epsilon}\rp)\rp]  ~,
	\label{e:Ifirst}
 \\
   I_{s,\,I}(K_*) &=&  \int_0^1 
    	 d\tau \frac{\sqrt{0.034}~\tau^2}{\lp(\tau+\frac{2v_L}{\epsilon}\rp) 
	 \sqrt{1+\lp(\frac{K_*\tau^{2/3}}{12}\rp)^4+
    \lp(\frac{K_*\tau^{2/3}}{6}\rp)^{7/2}}}  	
	 \times \nn \\ && 
\sin\lp[2\pi  K_*\lp(\frac{\tau}{2v_L} +\frac{1}{\epsilon}\rp)\rp]  ~,
\eea
while for $ t \geq t_*$ the integrals can be written as
\bea
   I_{c,\,II}(K_*) &=&     \int_1^{\frac{t_{\rm fin}}{t_L^*}} 
 	 d\tau \frac{\sqrt{0.034}~\tau^{1/3}}{\lp(\tau+\frac{2v_L}{\epsilon}\rp) 
       \sqrt{1+\lp(\frac{K_*\tau^{2/3}}{12}\rp)^4+
      \lp(\frac{K_*\tau^{2/3}}{6}\rp)^{7/2}}}  	
	 \times \nn \\ && 
\cos\lp[2\pi  K_*\lp(\frac{\tau}{v_L} +\frac{1}{\epsilon}\rp)\rp]  ~, \label{dominantEW}
	\\
  I_{s,\,II}(K_*) &=&  \int_1^{\frac{t_{\rm fin}}{t_L^*}}
   	 d\tau \frac{\sqrt{0.034}~\tau^{1/3}}{\lp(\tau+\frac{2v_L}{\epsilon}\rp) 
	 \sqrt{1+\lp(\frac{K_*\tau^{2/3}}{12}\rp)^4+
   \lp(\frac{K_*\tau^{2/3}}{6}\rp)^{7/2}}}  	
	 \times \nn \\ && 
	\sin\lp[2\pi K_*\lp(\frac{\tau}{2v_L} +
         \frac{1}{\epsilon}\rp)\rp]  ~.
\eea

Secondly we consider a red magnetic field with $n=-1.8<-3/2$, that implies $A=-3/5$, $B=29/10$ and $C=1.4$. 
In this case the four integrals  read for $t_{\rm in}  \leq t < t_*$ 
\bea
   I_{c,\,I}(K_*) &=&     \int_0^1 
 	 d\tau \frac{\tau^{9/5} (K_*/40)^{-3/10} } {\lp(\tau+\frac{2v_L}{\epsilon}\rp) 
	 \sqrt{1+\lp(\frac{K_*\tau^{2/3}}{1.4}\rp)^{29/10}}}  	
	 \times \nn \\ && 
\cos\lp[2\pi  K_*\lp(\frac{\tau}{2v_L} +\frac{1}{\epsilon}\rp)\rp]  ~,
	\\
   I_{s,\,I}(K_*) &=&  \int_0^1 
    	 d\tau \frac{\tau^{9/5}(K_*/40)^{-3/10}}{\lp(\tau+\frac{2v_L}{\epsilon}\rp) 
	 \sqrt{1+\lp(\frac{K_*\tau^{2/3}}{1.4}\rp)^{29/10}}}  	
	 \times \nn \\ && 
\sin\lp[2\pi  K_*\lp(\frac{\tau}{2v_L} +\frac{1}{\epsilon}\rp)\rp]  ~,
\eea
while for $ t \geq t_*$ the integrals can be written as
\bea
   I_{c,\,II}(K_*) &=&     \int_1^{\frac{t_{\rm fin}}{t_L^*}} 
  d\tau \frac{\tau^{2/15}(K_*/40)^{-3/10}}{\lp(\tau+\frac{2v_L}{\epsilon}\rp) 
 \sqrt{1+\lp(\frac{K_*\tau^{2/3}}{1.4}\rp)^{29/10}}}  	
	 \times \nn \\ && 
\cos\lp[2\pi  K_*\lp(\frac{\tau}{2v_L} +\frac{1}{\epsilon}\rp)\rp]  ~, \label{dominantINF}
	\\
  I_{s,\,II}(K_*) &=&  \int_1^{\frac{t_{\rm fin}}{t_L^*}} 
  d\tau \frac{\tau^{2/15}(K_*/40)^{-3/10}}{\lp(\tau+\frac{2v_L}{\epsilon}\rp) 
	 \sqrt{1+\lp(\frac{K_*\tau^{2/3}}{1.4}\rp)^{29/10}}}  	
	 \times \nn \\ && 
\sin\lp[2\pi K_*\lp(\frac{\tau}{2v_L} +\frac{1}{\epsilon}\rp)\rp]  ~. 
\label{e:Ilast}
\eea
Inserting typical values for the above quantities we perform a numerical integration, and we find that the first phase, the 'switching on' of the 
inverse cascade, is completely irrelevant for the final result for most of
the spectrum. It does, however affect the peak position and the decay law as
we discuss in Section~\ref{s:peak}. The numerical solutions of the integrals are shown in Figs.~\ref{fig:GWK} and \ref{fig:redGWK}. 

\subsection{The fits for the GW spectrum}
\label{app:fits}

In deriving the analytical fits to the numerical GW spectra, 
Eqs.~(\ref{OmGWEW}) and (\ref{OmGWINF}), we have been guided by analytic 
intuition of the behaviour of the integrals given in Eqs.~(\ref{e:Ifirst}) 
to (\ref{e:Ilast}) above. Here we give some details for the understanding of 
the fits. 

Let us start with the causal case, $n=2$. First of all, the main contribution to the GW spectrum comes from the integral in Eq.~(\ref{dominantEW}), {\it i.e.} the cosine part in (\ref{drhoGW}). For very small values of $K_*$, below the characteristic wave number $k \leq 1 /t_{\rm fin}$, the cosine does not oscillate: therefore, we expect to inherit directly the slope of the anisotropic stress. For the causal case this is flat, consequently we expect a $K_*^3$ behaviour, coming from (\ref{app:rhoGW}). The constant $\epsilon_1$ is fixed by the large wavelength limit of $\mathcal{I}_{\rm GW}(K_*)$, given mainly by the integral of Eq.~(\ref{dominantEW}) evaluated at the upper boundary $t_{\rm fin}$:
\ben
   \mathcal{I}_{\rm GW}(K_*\to 0) \simeq 7.73 \,\ti\Pi(0) \lp(\frac{T_*}{T_{\rm fin}}\frac{2v_L}{\ep}\rp)^{2/3} K_*^3\equiv \ep_1\,K_*^3~.
\een
For higher values of the wave number, the main contribution to the integral comes roughly from the first oscillation of the cosine in Eq.~(\ref{dominantEW}) (note that the integrand decays with time). This can be accounted for by integrating only up to the time $t\simeq 1/k$, causing a change of slope of the GW spectrum, which now results in $\mathcal{I}_{\rm GW} (K_*) \propto K_*^{7/3}$. In the main text this slope is set to $K_*^2$, which corresponds to the best fit result from the numerical evaluation of the integral (see~Fig.~\ref{fig:GWK}). These analytical considerations are in fact quite crude and lead to slopes which are not very precise. The parameter $\epsilon_2$  is determined by the matching at the limiting value $k=1/t_{\rm fin}$, which is the value of the wave number for which the cosine starts to oscillate: 
$$\ep_2 \simeq 0.07 \lp(  v_L^2 \,\ep\, \frac{T_{\rm fin}}{T_*}\rp)^{1/3}\,.$$
This behaviour continues until $k$ becomes of the order of $1/t_*$. Above this 
value, the time dependence of the integrand is no longer $\tau^{-2/3}$ but 
$\tau^{1/3}$ (see~Eq.~(\ref{dominantEW})). This results in a further change 
in the slope of the spectrum, which now becomes $\mathcal{I}_{\rm GW} (K_*) 
\propto K_*^{1/3}$. In the main text this slope is set to $\sqrt{K_*}$, 
again according to the numerical evaluation of the integral. By continuity, 
the paramter $\ep_3$ in Eq.~(\ref{OmGWEW}) is
$$  \ep_3 \simeq 4\cdot 10^{-3} \lp( v_L^2\,\ep^{11/2}\,
     \frac{T_{\rm fin}}{T_*}\rp)^{1/3} \,.  $$

In the inflationary case, the main contribution to the GW spectrum comes again
from the integral in Eq.~(\ref{dominantINF}). For very small values of $K_*$, below the characteristic wave number $k \leq 1 /t_{\rm fin}$, we expect to inherit the slope of the anisotropic stress. The constant $\epsilon_4$ is given by the large wavelength limit of $\mathcal{I}_{\rm GW}(K_*)$:
\be
\mathcal{I}_{\rm GW}(K_*\to 0)\simeq 68\,\ti\Pi(K_*)\left(\frac{T_*}{T_{\rm fin}}\right)^{4/15} K_*^3=\ep_4\,K_*^{2n+6}~.
\ee
The above formula is valid for $n=-1.8$. For higher values of the wave number, the same argument as in the causal case applies, and we integrate only up to $t\simeq 1/k$: the slope in wave number of the GW spectrum now results in $\mathcal{I}_{\rm GW} (K_*) \propto K_*^{(2n+10)/3}$. By continuity, we obtain (again for $n=-1.8$) $\ep_5 \simeq 619/(2\pi)^{4/15}$. This behaviour continues until the wave number for which the final time of turbulence $t_{\rm fin}(k)$, given in Eq.~(\ref{e:tfin}), becomes smaller than $1/k$: this happens for $K_*\simeq ((2\pi)^5/R_*^9)^{1/7}$, see~the first line of Eq.~(\ref{e:tk2inf}). For higher wave numbers, the upper limit of integration has a different $k-$behaviour which translates to the slope $\mathcal{I}_{\rm GW}(K_*) \propto K_*^{-(2+6n)/5}$. By continuity,
$\ep_6 \simeq 619/R_*^{12/25}\simeq 22$ for $n=-1.8$ and $R_* \simeq 10^{3}$ for inflation.

\end{document}